\documentclass[usegraphicx,useAMS,usenatbib]{mn2e}
\usepackage[T1]{fontenc}
\usepackage{verbatim}
\usepackage[normalem]{ulem} 
\usepackage{amsfonts,amsmath,amssymb} 
\usepackage{lipsum,multicol}
\usepackage{times}
\usepackage[italic]{mathastext} 
\usepackage{enumitem}
\usepackage[hidelinks]{hyperref}

\usepackage{multirow}

\usepackage{color}
\usepackage{upgreek}
\usepackage{flushend}
\usepackage{tikz}

\newcommand{\hinv}{h^{-1}}

\newcommand{\redmapper}{redMaPPer}

\newcommand{\msun}{\mathrm{M}_{\odot}}

\newcommand{\lnM}{{\rm ln}M}

\newcommand{\lamRM}{\lambda_{\rm RM}}

\newcommand{\chandra}{\emph{Chandra}}
\newcommand{\xmm}{\emph{XMM}}
\newcommand{\RM}{redMaPPer}

\defcitealias{Evrard:2014}{E14}

\newcommand{\rlambdat}{\ensuremath{r_{\lambda  T}}}
\newcommand{\rsqlambdat}{\ensuremath{r^2_{\lambda  T}}}
\newcommand{\varmulambda}{\ensuremath{\sigma^2_{\ln M \,|\,\lambda}}}
\newcommand{\sigmamulambda}{\ensuremath{\sigma_{\ln M \,|\,\lambda}}}

\newcommand{\sigmalnMlambda}{\ensuremath{\sigma_{\ln M \,|\,\lambda}}}
\newcommand{\sigmaTxlambda}{\ensuremath{\sigma_{\ln T \,|\,\lambda}}}
\newcommand{\varTxlambda}{\ensuremath{\sigma^2_{\ln T \,|\,\lambda}}}
\newcommand{\varmuTx}{\ensuremath{\sigma^2_{\ln M \,|\, T }}}
\newcommand{\sigmamuTx}{\ensuremath{\sigma_{\ln M \,|\, T}}}

\usepackage{todonotes}

\hypersetup{draft}

\title[Mass Variance of DES redMaPPer Clusters]{Mass Variance from Archival X-ray Properties of Dark Energy Survey Year-1 Galaxy Clusters }

\author[A.~Farahi, et al.]{
\parbox{\textwidth}{
\Large
\mbox{A.~Farahi,$^{1,2}$\thanks{corresponding author: \href{mailto:aryaf@umich.edu}{afarahi@andrew.cmu.edu}}}
X.~Chen,$^{2,3}$
A.~E.~Evrard,$^{4,2}$
D.~L.~Hollowood,$^{5}$
R.~Wilkinson,$^{6}$
S.~Bhargava,$^{6}$
P.~Giles,$^{6}$
A.~K.~Romer,$^{6}$
T.~Jeltema,$^{5}$
M.~Hilton,$^{7}$
A.~Bermeo,$^{6}$
J.~Mayers,$^{6}$
C.~Vergara Cervantes,$^{6}$
E.~Rozo,$^{8}$
E.~S.~Rykoff,$^{9,10}$
C.~Collins,$^{11}$
M.~Costanzi,$^{12}$
S.~Everett,$^{5}$
A.~R.~Liddle,$^{13,14,15}$
R.~G.~Mann,$^{14}$
A.~Mantz,$^{9}$
P.~Rooney,$^{6}$
M.~Sahlen,$^{16}$
J.~Stott,,$^{17}$
P.~T.~P.~Viana,$^{18}$
Y.~Zhang,$^{19}$
J.~Annis,$^{19}$
S.~Avila,$^{20}$
D.~Brooks,$^{21}$
E.~Buckley-Geer,$^{19}$
D.~L.~Burke,$^{9,10}$
A.~Carnero~Rosell,$^{22,23}$
M.~Carrasco~Kind,$^{24,25}$
J.~Carretero,$^{26}$
F.~J.~Castander,$^{27,28}$
L.~N.~da Costa,$^{23,29}$
J.~De~Vicente,$^{22}$
S.~Desai,$^{30}$
H.~T.~Diehl,$^{19}$
J.~P.~Dietrich,$^{31,32}$
P.~Doel,$^{21}$
B.~Flaugher,$^{19}$
P.~Fosalba,$^{27,28}$
J.~Frieman,$^{19,33}$
J.~Garc\'ia-Bellido,$^{34}$
E.~Gaztanaga,$^{27,28}$
D.~W.~Gerdes,$^{4,2}$
D.~Gruen,$^{35,9,10}$
R.~A.~Gruendl,$^{24,25}$
J.~Gschwend,$^{23,29}$
G.~Gutierrez,$^{19}$
K.~Honscheid,$^{36,37}$
D.~J.~James,$^{38}$
E.~Krause,$^{39}$
K.~Kuehn,$^{40}$
N.~Kuropatkin,$^{19}$
M.~Lima,$^{41,23}$
M.~A.~G.~Maia,$^{23,29}$
J.~L.~Marshall,$^{42}$
P.~Melchior,$^{43}$
F.~Menanteau,$^{24,25}$
R.~Miquel,$^{44,26}$
R.~L.~C.~Ogando,$^{23,29}$
A.~A.~Plazas,$^{43}$
E.~Sanchez,$^{22}$
V.~Scarpine,$^{19}$
M.~Schubnell,$^{2}$
S.~Serrano,$^{27,28}$
I.~Sevilla-Noarbe,$^{22}$
M.~Smith,$^{45}$
F.~Sobreira,$^{46,23}$
E.~Suchyta,$^{47}$
M.~E.~C.~Swanson,$^{25}$
G.~Tarle,$^{2}$
D.~Thomas,$^{20}$
D.~L.~Tucker,$^{19}$
V.~Vikram,$^{48}$
A.~R.~Walker,$^{49}$
and J.~Weller$^{31,50,12}$
\begin{center} (DES Collaboration) \end{center}
\bigskip
\small{{\em Author affiliations are listed at the end of this paper.}}
}
}

\begin{document}
\date{\today}
\pagerange{\pageref{firstpage}--\pageref{lastpage}} \pubyear{2018}
\maketitle
\label{firstpage}

\begin{abstract}
Using archival X-ray observations and a log-normal population model, we estimate constraints on the intrinsic scatter in halo mass at fixed optical richness for a galaxy cluster sample identified in Dark Energy Survey Year-One (DES-Y1) data with the redMaPPer algorithm.  We examine the scaling behavior of X-ray temperatures, $T_X$, with optical richness, $\lamRM$, for clusters in the redshift range $0.2<z<0.7$. X-ray temperatures are obtained from \chandra\ and \xmm\ observations for 58 and 110 \RM\ systems, respectively.  Despite non-uniform sky coverage, the $T_X$ measurements are $> 50\%$ complete for clusters with $\lamRM > 130$. Regression analysis on the two samples produces consistent posterior scaling parameters, from which we derive a combined constraint on the residual scatter, $\sigmaTxlambda = 0.275 \pm 0.019$.  Joined with  constraints for $T_X$ scaling with halo mass from the Weighing the Giants program and richness--temperature covariance estimates from the LoCuSS sample, we derive the richness-conditioned scatter in mass, $\sigmalnMlambda = 0.30 \pm 0.04\, _{({\rm stat})} \pm 0.09\, _{({\rm sys})}$, at an optical richness of approximately 70.  Uncertainties in external parameters, particularly the slope and variance of the $T_X$--mass relation and the covariance of $T_X$ and $\lamRM$ at fixed mass, dominate the systematic error. The $95\%$ confidence region from joint sample analysis is relatively broad, $\sigmalnMlambda \in [0.14, \, 0.55]$, or a factor ten in variance.
\end{abstract}


\section{Introduction} 
\label{sec:introduction}

Population statistics of galaxy clusters are acknowledged as a valuable probe of cosmological parameters \citep{Allen:2011review,Weinberg:2013,Huterer:2018}, as illustrated by analysis of modern cluster samples \citep[\textsl{e.g.},][]{Vikhlinin:2009, Rozo:2010, Benson:2013, Mantz:2014, deHaan:2016, Bocquet:2019}, and anticipated from larger and deeper cluster samples being assembled. The Dark Energy Survey \citep[DES,][]{DES:0verview} is  identifying clusters using color-based searches in five-band optical photometry. A small initial sample from the Science Verification survey phase, with 786 clusters, \citep{Rykoff:2016} is supplemented by a Year-1 (Y1) data sample containing $\sim 7,000$ clusters with 20 or more statistical galaxy members \citep{DESY1-WL:2018}.

The population statistics approach relies on comparing the number and spatial clustering of galaxy clusters, as a function of their observable properties and redshift, to theoretical expectations derived from simulations of dark matter halos, particularly the Halo Mass Function (HMF) \citep[e.g.,][]{Jenkins:2001, Evrard:2002, Tinker:2008, Murray:2013}. To connect halo and cluster properties, a probabilistic model commonly referred to as the mass--observable relation is employed to map host halo mass to multiple cluster observables. This paper focuses on the statistical relationships between optical and X-ray properties of a cluster and the underlying total mass of the halo hosting it.

Ensemble-averaging, or stacking, to estimate mean mass as a function of galaxy richness has been applied to the DES-Y1 cluster sample by \citet{DESY1-WL:2018}.  The process of stacking has the drawback that it integrates out the variance in halo mass, $M$, conditioned on galaxy richness, $\lambda$.  A complementary inference technique is needed to determine the width and shape of the conditional probability distribution, $\Pr(M \, | \, \lambda)$.

Both observations \citep{Pratt:2009, Reichert:2011, Mahdavi:2013, Lieu:2016xxl, Mantz:2016-relaxedIII, Mantz:2016WtG-V} and simulations \citep[][]{Evrard:2008, Stanek:2010, Farahi:2017} support a log-normal form for observable-mass conditional distributions.  This form, coupled with a low-order polynomial approximation for the HMF, yields analytic expressions for the space density as a function of multiple observable properties as well as property-conditioned statistics of the massive halos hosting groups and clusters \citep[][hereafter, E14]{Evrard:2014}.  We employ this model in our analysis, with particular emphasis on conditional property covariance.

The red-sequence Matched-filter Probabilistic Percolation (\RM) identifies clusters using an empirically-calibrated, matched-filter model for old, red galaxies \citep{Rykoff:2014}.  The algorithm outputs a probabilistic estimate of optical richness -- the count of red galaxies inside a cluster -- along with a mean cluster redshift and a set of up to five likely central galaxies.  Previous studies found this photometric cluster-finder algorithm produces a highly complete and pure cluster sample with accurate redshift estimates \citep{Rykoff:2012, Rozo:2014, Rozo:2015, Rozo:2015IV}.

The Sloan Digital Sky Survey (SDSS) DR-8 \RM\ cluster sample \citep{Rykoff:2014} has recently been combined with ensemble-average weak lensing masses \citep{Simet:2017} to produce cosmological constraints \citep{SDSS:2018cosmology}. A similar analysis is underway for DES-Y1 \citep{DESY1-WL:2018}.  In both of these works, marginalization over the weakly constrained scatter between mass and richness weakens posterior likelihoods of cosmological parameters.  The aim of our work is to provide an empirical constraint on the mass-richness variance, a result that will be combined with other systematics calibration effort to refine and improve likelihood analysis of cluster counts for cosmology.

Integrated measures of clusters such as \RM\ richness, $\lamRM$, X-ray temperature, $T_X$, and luminosity, $L_X$, are proxies for host halo mass in that each scales as a (typically) positive power of $M$.  In general, each proxy has intrinsic variance generated by internal dynamics within halos, as well as extrinsic scatter caused by projection, measurement uncertainties and other effects. 
For the intrinsic component, the log-normal property covariance model of \citetalias{Evrard:2014} provides expressions that link proxy properties to each other and to unobservable host halo mass.  The expressions involve the local slope and curvature of the HMF because of the convolution required to map mass to the observed measures. 

Here we study the scaling behavior of $T_X$ as a function of $\lamRM$ for a \RM\ sample of clusters identified in DES-Y1 imaging data within the redshift range, $z \in [0.2, 0.7]$ \citep{DESY1-WL:2018}. X-ray properties of clusters contained in archival \chandra\ or \xmm\ pointings are measured via the \textbf{M}ass \textbf{A}nalysis \textbf{T}ool for \textbf{Cha}ndra \citep[MATCha,][]{Hollowood:2018} or XCS data analysis pipelines (Giles et al. in preparation), respectively.  We employ the Bayesian regression model of \citet{Kelly2007} to estimate parameters of the conditional scaling, $\Pr(\ln T_X \, | \, \lamRM)$. 

The inference of mass scatter requires additional information, namely the $T_X$--$M_{\rm wl}$ scaling relation and the $\lamRM$--$T_X$ covariance at fixed halo mass. These additional quantities are taken from previous studies \citep[][Farahi et al. in prep.]{Mantz:2016WtG-V}.  Uncertainties on the inferred scatter are determined by marginalizing over uncertainties in the model priors. 

The structure of this paper is as follows. In Section \ref{sec:desy1}, we introduce the cluster sample and X-ray follow-up programs of the optically-selected clusters. In Section \ref{sec:population}, we describe the regression algorithm and the population model employed to obtain an estimate of the mass--richness scatter, with results presented in Section \ref{sec:results}. In Section \ref{sec:discussion}, we discuss our treatment of systematic uncertainties. Finally, we conclude in Section \ref{sec:conclusion}. Appendix \ref{app:cluster-cat} provides the tables of cluster properties employed in this work. Appendix \ref{app:centering-sensitivity} provides corrections for a small number of richness measurements using the X-ray emission peak locations of \chandra\ and \xmm\ observations. Finally in Appendices \ref{app:correlation} and \ref{app:running}, we present the richness--temperature correlation at fixed halo mass and upper limits on the running of temperature variance at fixed optical--richness, respectively.

We assume a flat $\Lambda$CDM cosmology with $\Omega_{\rm m}=0.3$ and $H_0=70$ km s$^{-1}$ Mpc$^{-1}$. Distances and masses, unless otherwise noted, are defined as physical quantities with this choice of cosmology, rather than in comoving coordinates. We denote the mass inside spheres around the cluster center as $M_{200\rm{c}}$, corresponding to an overdensity of 200 times the critical matter density at the cluster redshift.


\section{DES-Y1 data}
\label{sec:desy1}

This work is based on data obtained during the DES-Y1 observational season, between 31$^{\rm st}$, August 2013 and 9$^{\rm th}$, February 2014 \citep{Y1gold}. During this period 1839 deg$^2$ was mapped out in three to four tilings using $g,r,i,z$ filters. This strategy produces a shallower survey depth compared to the full-depth Science Verification data, but it covers a significantly larger area.  We use approximately 1,500 deg$^2$ of the main survey split into two contiguous areas, one overlapping the South Pole Telescope (SPT) Sunyaev-Zel'dovich Survey area, and the other overlapping the Stripe-82 (S82) deep field of SDSS. The sky footprint is illustrated in Fig. 1 of \citet{DESY1-WL:2018}.

We first describe the main data products used in our analysis and refer the reader to corresponding papers for a more detailed overview.  Imaging and galaxy catalogs associated with the \RM\ catalog used here are publicly available\footnote{\url{https://des.ncsa.illinois.edu/releases/dr1}} in the first DES data release \citep[DES DR-1,][]{DES-DR1:2018} and the Y1A1 GOLD wide-area object catalog \citep{Y1gold}.

\subsection{Optical cluster catalog}
\label{sec:redmapper}

 We employ a volume-limited sample of galaxy clusters detected in the DES-Y1 photometric data using version 6.4.17 of the \RM\ cluster-finding algorithm \citep{Rykoff:2016}.  The \RM\ algorithm identifies clusters of red-sequence galaxies in the multi-dimensional space of four-band magnitudes and sky position. Starting from an initial spectroscopic seed sample of galaxies, the algorithm iteratively fits a model for the local red-sequence.  It then performs a matched filter step to find cluster candidates and assign membership probabilities to potential members.  Starting with a most likely central galaxy, ideally the brightest cluster galaxy, member weights, $p_{\rm mem}$, of additional cluster galaxies are computed with a matched-filter algorithm based on spatial, color, and magnitude filters \citep{Rykoff:2014}.  The method is iterative, and the ambiguity of selecting a central galaxy is recognized by recording likelihoods for up to five central galaxies in each cluster.  The final richness, $\lamRM$, is defined as the sum of the $p_{\rm mem}$ values of its member galaxies.  

\begin{figure*}
    \centering
    \includegraphics[width=0.48\textwidth]{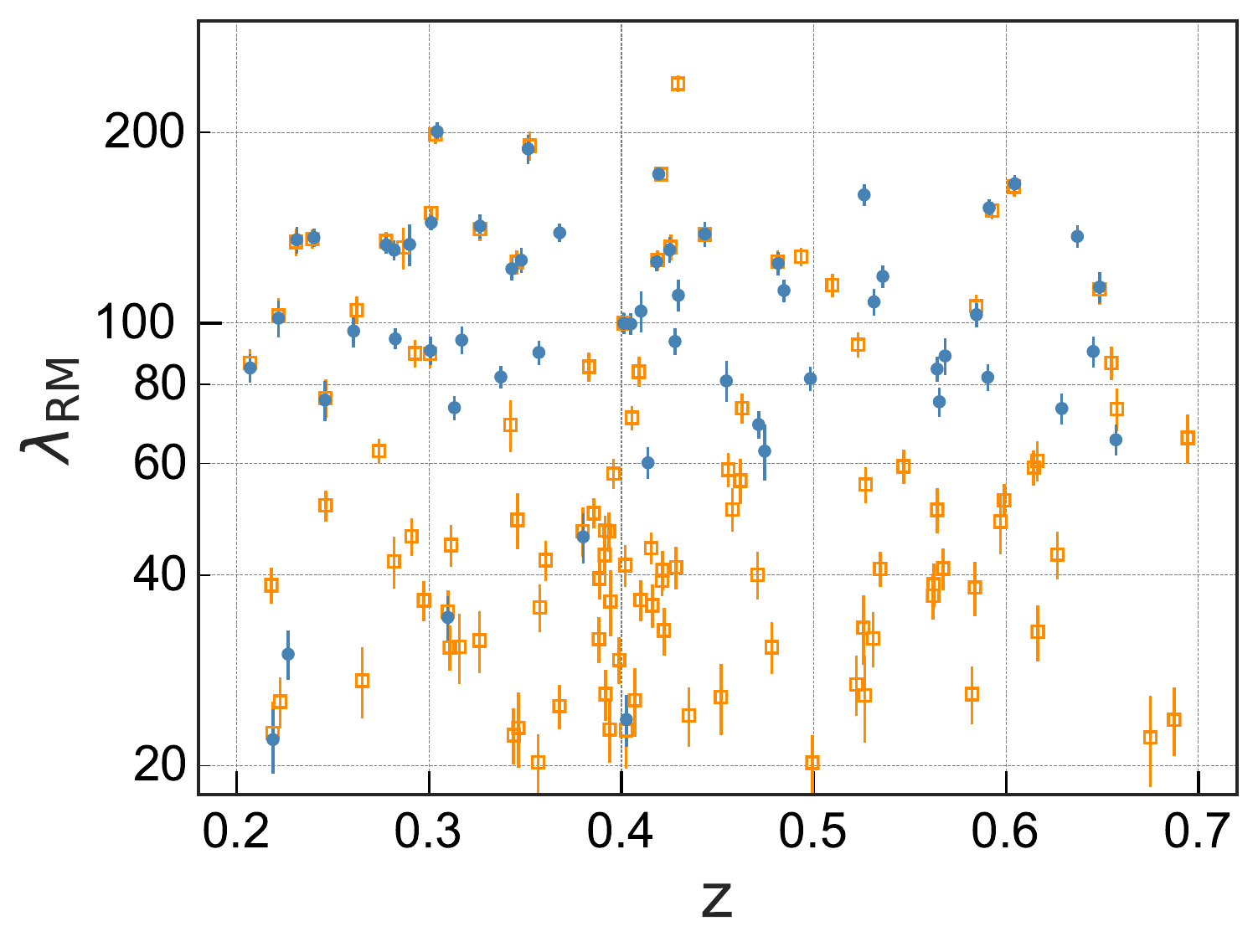}
    \includegraphics[width=0.48\textwidth]{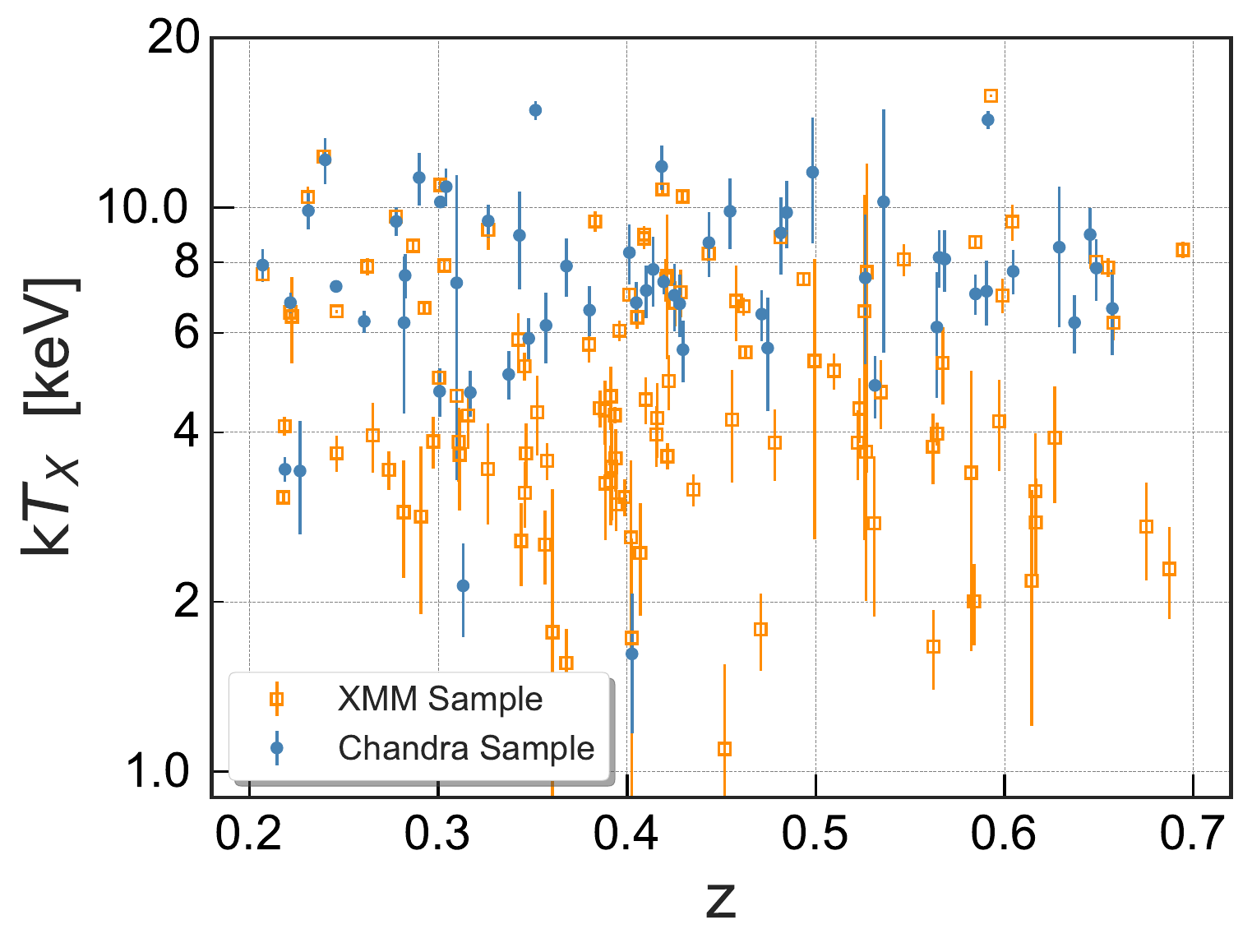}
    \caption{Supplementary sample properties.  Left and right panels show the  richness and hot gas temperatures as a function of cluster redshift obtained from \xmm\ and \chandra\ archival data. The optical-richness is re-measured at the location of the X-ray emission peak (see Section~\ref{sec:Xraycats} for more detail). \xmm\ temperatures are scaled to the \chandra\ according to equation~\ref{eq:temp-scale}. The error bars are $68\%$ measurement errors.}  
    \label{fig:data-summary}
\end{figure*}

\subsection{Supplemental X-ray catalogs}
\label{sec:Xraycats}

Table~\ref{tab:data-summary} summarizes contents of the \xmm\ and \chandra\ samples employed in this work, and Fig. \ref{fig:data-summary} shows the distribution of cluster samples as a function of their observables. The X-ray catalogs are provided in Appendix \ref{app:cluster-cat} and will be available from the online journal in machine-readable format. In the following, we detail how the \RM\ clusters are matched to the X-ray sources identified in \chandra\ and \xmm\ archival data, and how the X-ray properties of the matched sources are measured. 

The two methods produce independent luminosity and temperature estimates, and we adjust the latter to remove the known spectral bias between the two X-ray telescopes \citep{Schellenberger:2015}. The two catalogs have similar depth (median redshifts of $0.41$) while differing in their coverage of halo mass scale, reflected in Table~\ref{tab:data-summary} by offsets in the median values of mass proxies.  The median richness is 76 for \chandra, 38 for \xmm, and the respective median X-ray temperatures are $7.45$ and $4.41$~keV.  In terms of natural logarithms, these offsets are $0.48$ and $0.52$, respectively.  Both samples have range of a factor of ten in both $\lamRM$ and $T_X$ dimensions. 

We are concerned about the relation between the properties of the \RM-selected cluster observables and its host halo. Therefore, we need to correct for the fraction of the mis-centered population. Instead of modeling the mis-centered population, we correct our cluster observables with an associated X-ray center and re-estimate the optical--richness at X-ray peak (see Appendix~\ref{app:centering-sensitivity} for more detail). In the following, richness, $\lamRM$, implies the optical--richness assigned by the \RM\ algorithm at the X-ray peak, unless otherwise mentioned.

\begin{table}
\begin{center}
\caption{Supplemental X-ray Sample Size and Median Characteristics.} \label{tab:data-summary}
\begin{tabular}{|l|c|c|c|c|}
\hline
Source & $N_{\rm sam}$ & $z_{\rm med}$  & $\lambda_{\rm med}$ & ${\rm k}T_{X, med}$ [keV]  \\ \hline
Chandra & 58  & 0.41  & 76  & 7.45  \\ 
XMM   & 110 & 0.41  & 38  & 4.41 \\ \hline
\end{tabular}
\end{center}
\end{table}

\subsubsection{Chandra-\RM\ Catalogs}

The analysis of \chandra\ observations was conducted with the MATCha pipeline described in \citet[][]{Hollowood:2018}. We briefly outline the steps here. Starting from the volume-limited, $\lambda \geq 20$ \RM\ catalog, we analyze all archival \chandra\ data, public at the time of the analysis, which overlapped \RM\ cluster positions.  In brief, after standard data reduction and cleaning, we search for a significant X-ray cluster detection starting from the \RM\ position and iteratively re-centering toward the X-ray peak using an initial 500 kpc aperture.  If the cluster is X-ray detected (SNR$>5$), a spectrum is extracted, and we attempt to fit for the X-ray temperature, $T_X$.  An iterative process is employed to center, determine cluster temperature and luminosity in the same X-ray band, and estimate cluster radius based on the $T_X$ fit. To evaluate $T_X$, the metal abundance is fixed at 0.3$Z_{\odot}$, using the model from \citet{Anders:1989}. For clusters with sufficiently well-sampled data, the output of the MATCha algorithm includes the centroid location, $L_X$, and $T_X$ within a series of apertures, 500 kpc, $r_{2500}$, $r_{500}$, and core-cropped $r_{500}$. In this work, we only use core included $r_{2500}$ $T_X$ values. 

In addition, we estimate the X-ray emission peak position of each detected cluster for use in studying the \RM\ centering distribution \citep[][]{Hollowood:2018, Zhang:2018centering}.  The peak is determined, after smoothing the point-source subtracted cluster image with a Gaussian of 50 kpc width, as the brightest pixel within 500 kpc of \RM\ position.  We perform a visual check and then remove clusters for which the position, source spectrum, or background spectrum were significantly affected by instrumental chip edges or where the identified X-ray cluster was a foreground or background cluster not matched to the \RM\ cluster.

\subsubsection{XCS-\RM\ Catalogs}
\label{sec:xmm}

For the the \xmm-\RM\ analysis (Giles et al. in preparation), the \RM\ sample is matched to all {\em XMM} ObsIDs (with useable EPIC science data) under the requirement that the \RM\ position be within 13$^{\prime}$ of the aim point of the ObsID.  Next, the {\em XMM} observations were filtered based upon
exposure time. The exposure time is determined within a radius of 5 pixels centred on the \RM\ position, with the mean and median required to be >3ks and >1.5ks respectively. Here the mean is taken to be the exposure time averaged over the sum of each pixel, while the median refers to 50 percent of the pixels in the enclosed region. These cuts are applied to ensure the \RM\ cluster of interest is within the {\em XMM} FOV and has a sufficiently long exposure time for reliable SNR and $T_X$ measurements.

X-ray sources for each ObsID were then detected using the XCS Automated Pipeline Algorithm \citep[XAPA,][]{LloydDavies:2011}.  At the position of the most likely central galaxy of each \RM\ cluster, we match to all XAPA-defined extended sources within a comoving distance of 2~Mpc.  Cutout DES and {\em XMM} images are then produced and visually examined to assign a XAPA source to the optical cluster.  Through this process, the final \xmm-\RM\ sample contains 110 clusters. 

The luminosities and temperatures for the \xmm-\RM\ sample are derived using the {\em XCS} Post Processing Pipeline \cite[][]{LloydDavies:2011}, with updates presented in Giles et al. (in preparation).  Cluster spectra are extracted and fit in the 0.3 -- 7.9 keV band with an absorbed MeKaL model \citep{Liedahl:1995}.  The fits are performed using the {\tt xspec} package \citep{Arnaud96}, with the metallicity fixed at 0.3$\,Z_{\odot}$. Using an iterative procedure, spectra are extracted within $r_{2500}$.  We estimate an initial temperature within the XAPA source detection region \citep{LloydDavies:2011}, and an initial $r_{2500}$ estimated from the $r_{2500}-{\rm k}T$ relation of \cite{Arnaud05}.  A temperature is estimated within this $r_{2500}$, and hence an updated $r_{2500}$ estimated as above.  This process is then iterated until $r_{2500}$ converges to 10\%.  To asses the reliability of temperature estimates, variance of the temperature is calculated for each iteration. This involves generating a grid of 5-by-5 pixels and estimating the temperature for each region. We assign a mean temperature to each cluster which satisfies $\sigma (T_{x})/\langle T_{x} \rangle ) \le 0.25$, where $\sigma$ is the standard deviation and $\langle T_{x} \rangle$ is the mean of estimated temperatures. Similar to the \chandra\ analysis, the peak is determined, after smoothing the point-source subtracted cluster image with a Gaussian of 50 kpc width, as the brightest pixel within 500 kpc of \RM\ position.

The different X-ray detection method, combined with the larger collecting area of \xmm\ compared to the \chandra\ observatory, produces an X-ray sample for \xmm\ that both is larger and extends to lower richness than the \chandra\ detections.  We defer a detailed analysis of the X-ray selection processes used here to future work.  Here, we first analyze each sample independently, combining them after demonstrating consistency of posterior scaling parameters.

\subsubsection{X-ray temperature as primary mass proxy}

While X-ray luminosities, $L_X$, are measured for a larger number of clusters than are temperatures, the larger variance in non-core excised $L_X$ \citep{Fabian:1994, Mantz:2016WtG-V} and the complexities of modeling the supplemental survey masks motivate the choice of $T_X$ as the primary link to halo mass.  As we will see, systematic uncertainties limit the precision with which we can recover the scatter in underlying halo mass.

An important systematic effect that we address is the misalignment of X-ray cluster temperatures derived from the instruments on the \chandra\ and \xmm\ observatories \citep[][]{Schellenberger:2015}.  Since we are particularly interested in population variance, it is important to align the $T_X$ measurements before performing a joint sample regression. We use the calibration of \citet{Rykoff:2016} based on 41 SDSS \RM-selected clusters,
\begin{equation} \label{eq:temp-scale}
   \log_{10} ( T_X^{\rm Chandra} ) = 1.0133 \log_{10} ( T_X^{\rm XMM} ) + 0.1008\,,
\end{equation}
with temperatures in units of keV. \citet{Rykoff:2016} note that the above relation is consistent with that of \citet{Schellenberger:2015}. We employ the \chandra\ temperature scale in analysis below. 

Within our sample, there are $<20$ clusters with both \chandra\ and \xmm\ temperatures. The calibration relation from these clusters alone is consistent with that of \citet{Rykoff:2016}, but with larger uncertainties.   

$M_{\rm gas}$ is another low-scatter mass proxy \citep{Mulroy:2019}. Currently, $M_{\rm gas}$ measurement for these sets of clusters is unavailable. We are planning on employing $M_{\rm gas}$ measurement as another cluster mass proxy in a future work. 

\subsubsection{X-ray Completeness}

\begin{figure*}
    \centering
    \includegraphics[width=0.42\textwidth]{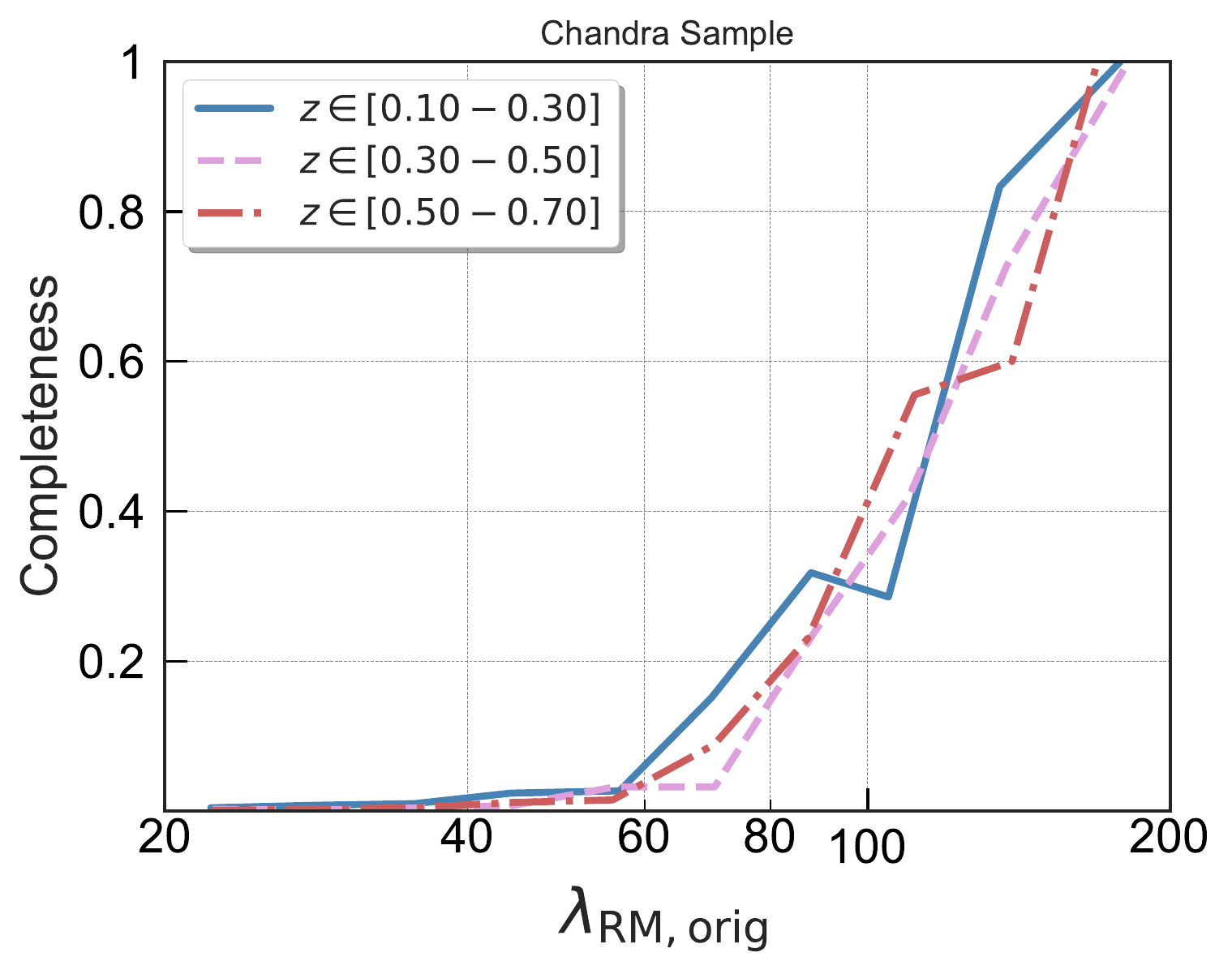} ~~~~~
    \includegraphics[width=0.42\textwidth]{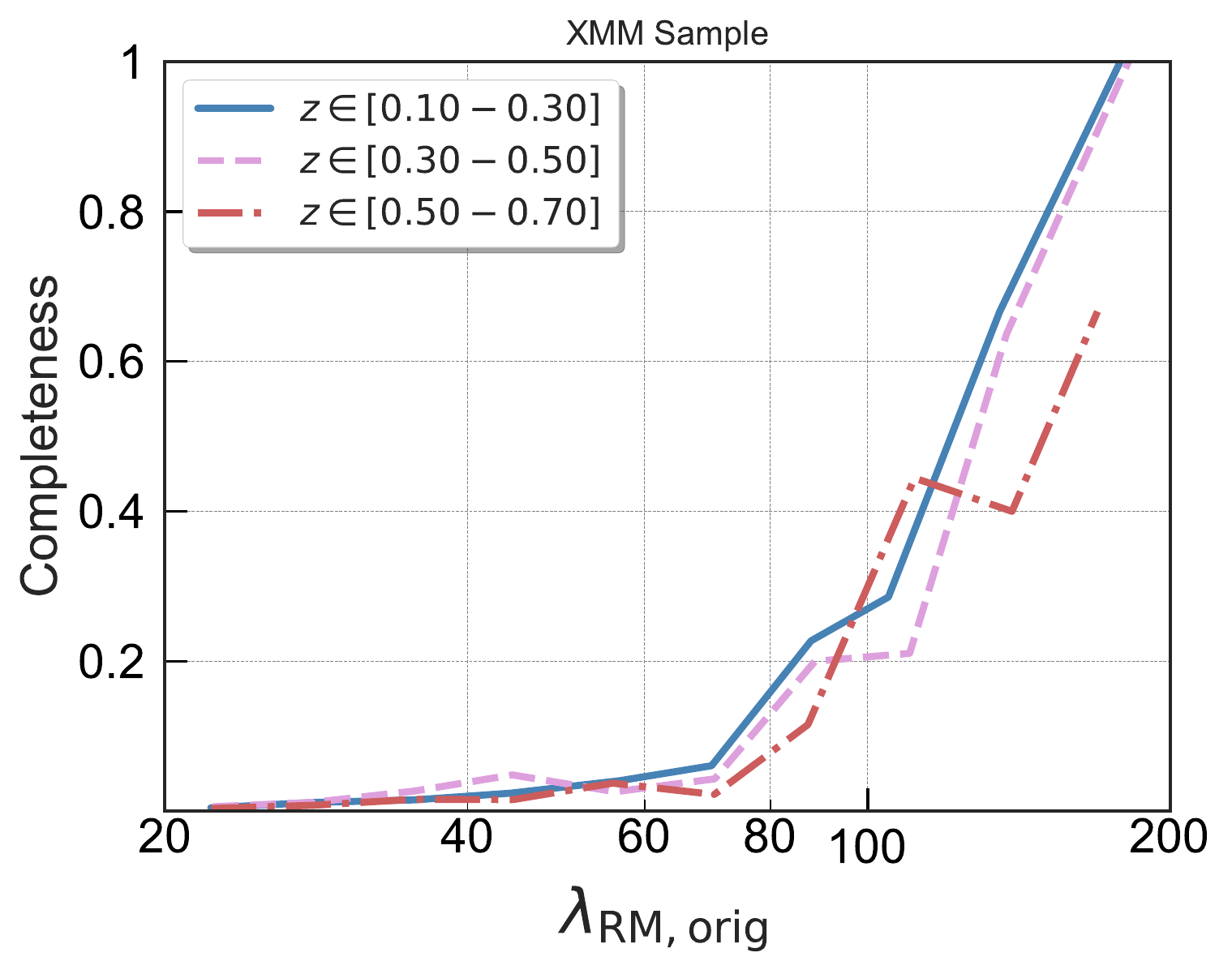}
    \caption{X-ray completeness of the supplementary samples.  Lines show the fractions of DES-Y1 \RM\ clusters with both $L_X$ and $T_X$ measurements from \chandra\ (left) and \xmm\ (right) archival data.  Solid, dashed, and dash-dot lines denoting increasing redshift bins given in the legend. Here, we employ the \RM\ original richness.
    }
    \label{fig:e-selection-function}
\end{figure*}

The supplemental samples, with fewer than 200 clusters, are far from complete relative to the full DES-Y1 \RM\ population of $\sim 7,000$ clusters.  The incompleteness is primarily due to the limited sky coverage of the two observatory archives to the depths required to detect distant clusters.  

If the X-ray signal-to-noise-ratio (SNR) is $>5$, typically a few hundreds of photons, there is enough signal to measure the X-ray luminosity; but at least 1,000 photons are needed to get a reliable estimation of the $T_X$. Fewer counts leads to larger errors, but not excluded from the sample. The variable depths of the archival pointings produce a complex pattern of flux limits across the optical sample. The observations also have different levels of background noise, adding to the complexity of X-ray selection modeling. Thorough synthetic observations \citep{Bahe:2012, ZuHone:2014} are needed to accurately model this selection function.  We defer such modeling to future work.  

The archival nature of the follow-up also produces a mix of previously known and newly detected X-ray systems. About $60\%$ of clusters are newly detected X-ray systems.  Fig. \ref{fig:e-selection-function} shows the fraction of \RM\ clusters with $T_X$ measurements as a function of \RM\ richness and redshift.  As may be expected, completeness is high for the largest clusters.  The sample is more than half complete in X-ray temperature at high optical richness values, $\lamRM \gtrsim 130$.  At lower richness, the completeness falls off, with more systems being found in the \xmm\ archival analysis.  We note in Appendix~\ref{app:running} that the posterior scaling parameters found in \S\ref{sec:results} are relatively insensitive to an imposed minimum richness threshold (see Fig.~\ref{fig:selection-sig}), but the effects of X-ray selection may affect our estimates of variance at low richness, particularly at high redshift. 

\section{Population Statistics} \label{sec:population}

The observed richness and X-ray temperature reflect properties of its host halo, subject to additional contributions from projected line-of-sight structure and other source of noise.  \citet{Costanzi:2018projection} develop a probabilistic model that maps intrinsic richness, $\lambda_{\rm true}$, to measured richness, $\lamRM$.  Generically, projection both widens the variance in $\Pr(\lamRM | M)$ and adds a moderate degree of skewness.  We do not apply corrections for projection effects, taking instead an approach that assumes $\Pr(\lamRM, T_X | M, z)$ is a bivariate log-normal.

The integrated stellar and gas mass fractions in halos extracted from recent hydrodynamic simulations follow a log-normal form, as validated at percent-level accuracy by \citet{Farahi:2017}. This form is also supported by previous cosmological simulations \citep{Evrard:2008, Stanek:2010, Truong:2018}.  Below, we show that normalized residuals in the measured scaling relation are consistent with a log-normal form, supporting this choice for our population inference model.

Our inference model has two steps: (i) $T_X$ is regressed against richness to determine scaling properties, particularly the residual\footnote{While the term ``intrinsic scatter'' is often used here, we use {\sl residual} so as to avoid confusion with the scatter associated with simulated halos and their properties measured within localized (typically spherical) cosmological volumes.  The residual variance in temperature at fixed observed cluster richness will be larger than the intrinsic variance at the relevant host halo richness because of projection.  Cluster members lying outside of the primary halo hosting the optical cluster add variance to the intrinsic relation and also bias the mean \citep{Costanzi:2018projection}.}  scatter in $T_X$ (Section \ref{sec:lam-t-scaling-realtion}), (ii) we combine the scatter of $T_X$--$\lamRM$ with weak-lensing mass--$\lamRM$ relations to infer the halo mass scatter at fixed optical richness (Section \ref{sec:halo-prop}).

\subsection{Regression model} \label{sec:regression}

We assume a log-normal form for the likelihood of a cluster of richness $\lambda$ to also have gas temperature $T$ 
 \begin{equation} \label{eq:ProbTxlambda}
\Pr (\ln T \, | \, \lambda, z) \  = 
 \frac{1}{ \sqrt{2 \pi} \sigmaTxlambda} 
 \exp\left\{ - \frac{ \left( \ln T - \langle \ln {\rm k} T \, | \, \lambda, z \rangle \right)^2}{ 2 \varTxlambda } \right\} .
\end{equation}
Here, and below, we may drop subscripts for simplicity of notation; $T \equiv T_X$ and $\lambda \equiv \lamRM$.  Following E14 notation, we write the log-mean scaling of ${\rm k} T$, expressed in keV, with richness as 
 \begin{equation} \label{eq:meanTxlambda}
 \langle \ln T \, | \, \lambda, z \rangle \ = \ 
 \left[ \pi_{T \,|\, \lambda} + 2/3 \ln(E(z)) \right] + \alpha_{T \,|\, \lambda} \, \ln(\lambda / \lambda_{\rm med}) , 
\end{equation}
where $\alpha_{T \,| \, \lambda}$ is the slope, $\lambda_{\rm med} = 70$ is the median richness of the joint sample, $\pi_{T \,| \, \lambda}$ is the logarithmic intercept at $z=0$, and $E(z)\equiv H(z)/H_{0}$ is the evolution of the Hubble parameter.  

We regress $T_X$ on $\lamRM$, rather than the other way around, because optical richness is the primary selection variable. Under the assumption that the X-ray temperature at fixed optical richness is either complete or randomly selected, explicit modeling of X-ray selection process is not required. We use the regression method of \citet[][]{Kelly2007}, which returns posterior estimates of the slope and normalization along with the residual variance, $\varTxlambda$.

The redshift dependence of the normalization in equation~(\ref{eq:meanTxlambda}) reflects a self-similar expectation, based on virial equilibrium, that $T \propto [E(z) M]^{2/3}$ \citep{Kaiser:1991, BryanNorman:1998}.  For observations spanning a range of redshift, the quantity $E^{-2/3}(z) {\rm k}T$ should be a closer reflection of halo mass, $M$, than temperature alone.  Over the redshift interval of our analysis, the $E^{-2/3}(z)$ factor decreases modestly, from $0.94$ at $z=0.2$ to $0.77$ at $z=0.7$. 

While the slope and scatter of scaling relations may be scale-dependent and/or evolving with redshift \citep[e.g.,][]{Farahi:2017,Ebrahimpour:2018}, our data are not yet rich enough to model these effects.  We crudely test richness dependence by splitting both samples into two non-overlapping samples, with different characteristic scales, at their pivot richness, and find no evidence of scale dependence in the posterior scaling relation parameters.

The regression method of \citet{Kelly2007} includes uncertainties associated with the independent variable, here $\ln \lamRM$, by assuming a mixture model in that variable.  The number of mixture elements is a free parameter in the method.  We use two components in our analysis, and have performed tests to demonstrate that our results are  insensitive to this hyperparameter.

\subsection{Mass scatter inference} \label{sec:massScatter}
In the \citetalias{Evrard:2014} population model, the variance in temperature of a sample conditioned on the selection variable, $\lambda$, is set by the joint (X-ray+optical) selection mass variance scaled by the slope, $\alpha_{T \,|\,M}$, of the temperature--mass relation, 
\begin{equation}
\varTxlambda =   
 \alpha^2_{T \,|\,M} \left[ \varmulambda + \varmuTx - 2 \rlambdat \, \sigmamulambda \, \sigmamuTx \ \right],
\label{eq:varTxlambda}
\end{equation} 
where $\varmulambda$ is the variance in halo mass at fixed optical richness (and similarly for temperature) and $\rlambdat$ is the correlation coefficient between log-richness and log-temperature at fixed halo mass.  We use the variance relationship,  $\varmuTx = \sigma^2_{\ln T \, |\, M} \, / \, \alpha^2_{T \, |\, M}$, and  assume that all parameters are constant with mass and redshift.

Rearranging the expression isolates what this study is after, the mass variance conditioned on optical richness,  
\begin{equation}
\varmulambda =  \varmuTx \, 
\left[ \left(\frac{\varTxlambda}{\sigma^2_{\ln T \,|\, M}} - (1-\rsqlambdat) \right)^{1/2} + \rlambdat  \right]^2  .
\label{eq:varmulambda}
\end{equation}
We ignore local curvature in the mass function, but note that its effect is to reduce the mass variance amplitude in equation~(\ref{eq:varmulambda}).  This suggests that the upper limits we derive below are somewhat conservative.  
 
If there is no property correlation, $\rlambdat = 0$, the above expression simplifies to 
\begin{equation}
\varmulambda =  \varmuTx \, 
 \left(\frac{\varTxlambda}{\sigma^2_{\ln T \,|\, M}} - 1 \right)  .
\label{eq:varmulambdasimple}
\end{equation} 
Note that the first term inside the parentheses is guaranteed to be greater than one because, when $r_{ab} = 0$, the simple Euclidean condition
\begin{equation}
\sigma^2_{b \, | \, a} = \sigma^2_{b \, | \, M }  + 
\left( \frac{\alpha_{b \, | \, M} }{\alpha_{a \, | \, M}} \right)^2  \sigma^2_{a \, | \, M } , 
\label{eq:var}
\end{equation} 
holds for any pair of properties $\{a,b\}$.

\section{Results} \label{sec:results}

In this section, we present temperature-richness scaling parameters derived from \chandra\ and \xmm\ data.  Consistent posterior constraints are found, motivating a joint analysis.  We then introduce additional priors on the missing elements of the residual mass variance conditioned on observed richness in \S~\ref{sec:halo-prop}, and present the resulting constraints. 

\begin{table}
\begin{center}
\caption{Best-fit parameters for the $T_X$--$\lamRM$ relation, equation~(\ref{eq:meanTxlambda}).}
\label{tab:TxlambdaParams}
\begin{tabular}{|l|c|c|c|}
\hline
  & Normalization & Slope & Residual scatter \\ 
Sample  & $e^{\pi_{T \,|\, \lambda}}$ [keV] & $\alpha_{T \, | \, \lambda }$ & $\sigma_{\ln T | \lambda}$ \\ \hline
Chandra & $5.23 \pm 0.26$      & $0.56 \pm 0.09$      & $0.260 \pm 0.032$  \\ 
XMM    & $4.88 \pm 0.15$      & $0.61 \pm 0.05$      & $0.289 \pm 0.025$   \\ \hline
Joint  & $4.97 \pm 0.12$      & $0.62 \pm 0.04$      & $0.275 \pm 0.019$   \\ \hline
\end{tabular}
\end{center}
\end{table}

\begin{figure*}
    \centering
    \includegraphics[width=0.34\textwidth]{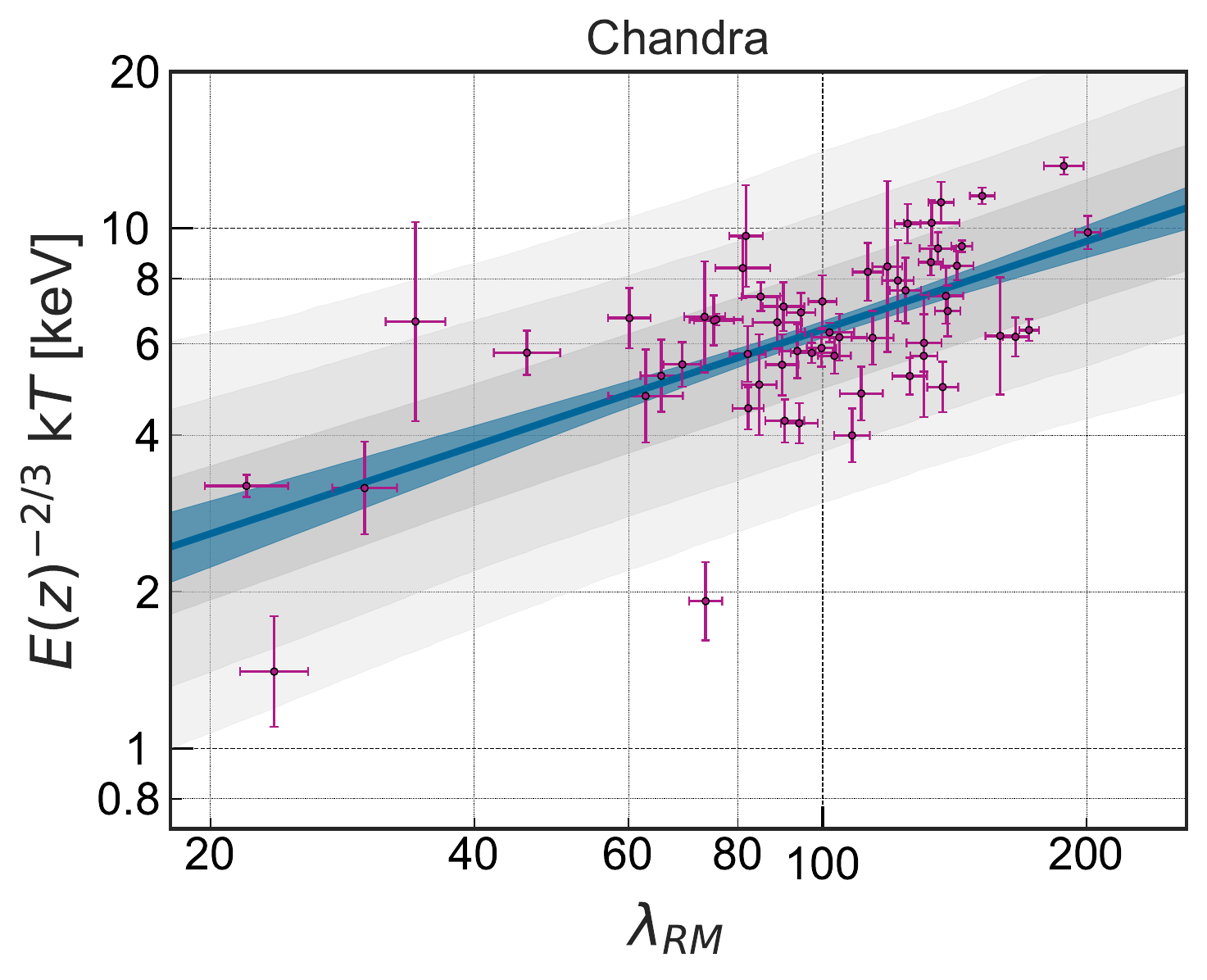}
    \includegraphics[width=0.32\textwidth]{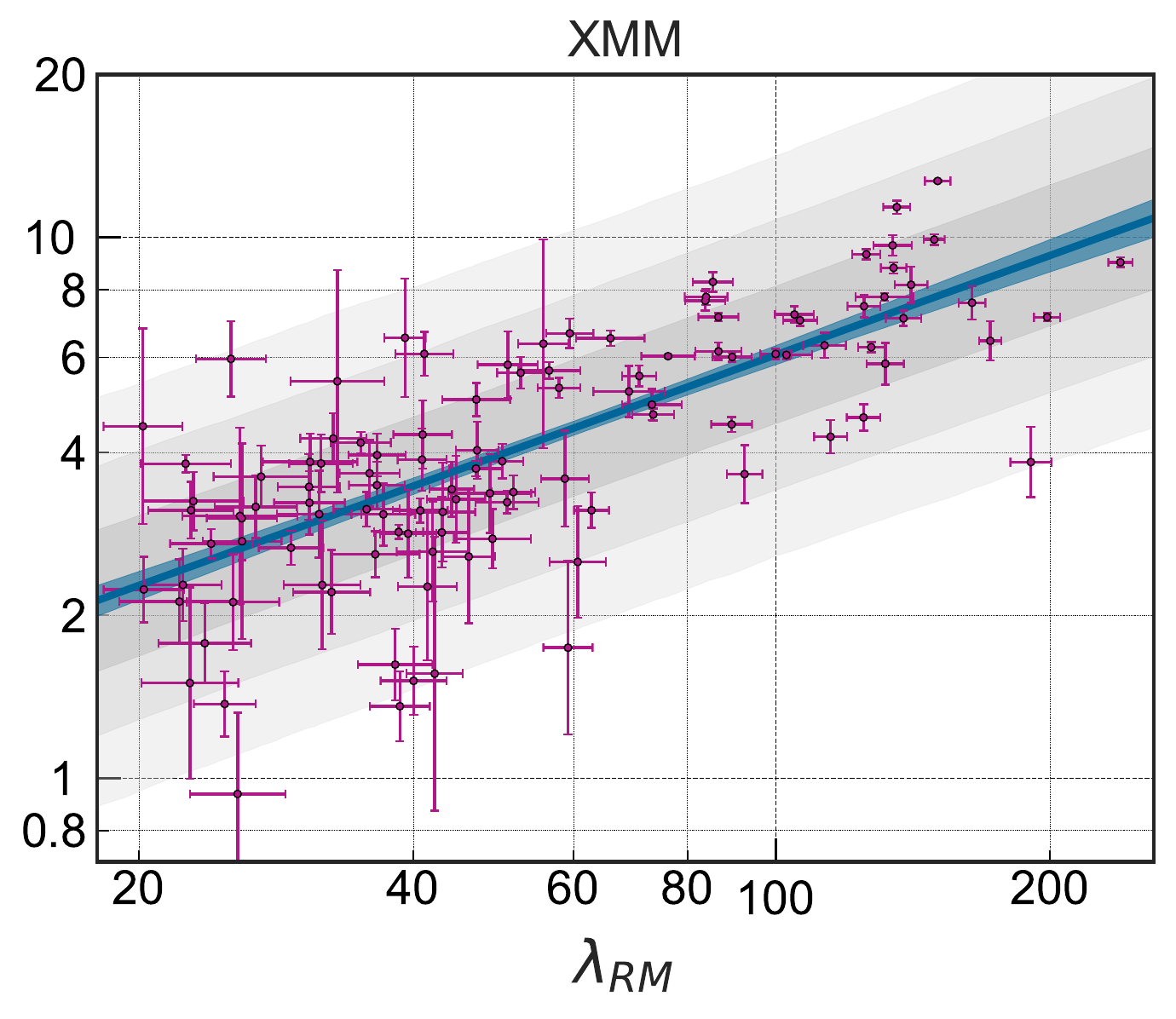} 
    \includegraphics[width=0.32\textwidth]{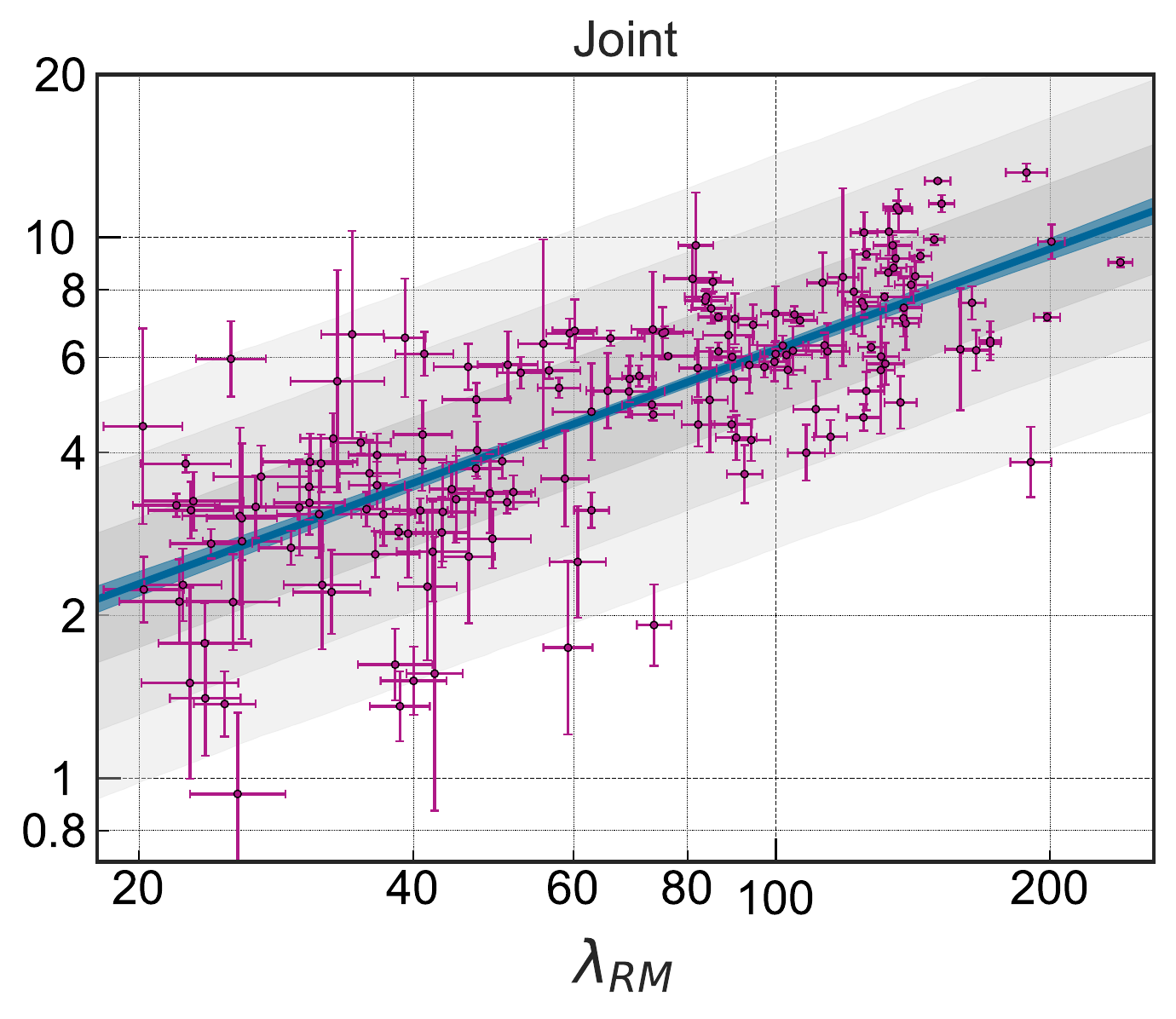}
    \caption{X-ray temperature--\RM\ richness scaling behavior from the \chandra, \xmm, and joint archival data samples (left to right) using a pivot richness of 70.  \xmm\ temperatures have been modified using equation~(\ref{eq:temp-scale}) to align with \chandra\ estimates.  In each panel, the blue line and blue shaded region are the best-fit and $68\%$ confidence interval of the mean logarithmic relation, equation~(\ref{eq:meanTxlambda}). Gray shaded regions show $1\sigma$, $2\sigma$, and $3\sigma$ residual scatter about the scaling relation. Fit parameters are given in Table~\ref{tab:TxlambdaParams}. Richness errors are provided directly by the \RM\ algorithm.}
    \label{fig:tx-lambda}
\end{figure*}

\subsection{\redmapper\ richness -- hot gas temperature relation} \label{sec:lam-t-scaling-realtion}

Figure \ref{fig:tx-lambda} shows the $T_X$--$\lamRM$ relation for the \chandra, \xmm, and joint samples respectively, and best-fit parameters are listed in Table \ref{tab:TxlambdaParams}.  The blue lines and shaded regions present best-fit and $68\%$ confidence intervals for the mean log scaling, equation~(\ref{eq:meanTxlambda}). In this regression, $\lamRM$ is remeasured at the location of the X-ray peak, for each cluster resulting in mostly small corrections and a small number of significant adjustments, as detailed in Appendix \ref{app:centering-sensitivity}.

We find consistent slopes of $0.56 \pm 0.09$ (\chandra) and $0.61 \pm 0.05$ (\xmm).  The \xmm\ temperature normalization, expressed in the \chandra\ system via the adjustment of equation~(\ref{eq:temp-scale}), is $4.88 \pm 0.15$~keV, roughly $1\sigma$ lower than the \chandra\ value of $5.23 \pm 0.26$. The \xmm\ temperature normalization, before the adjustment of equation~(\ref{eq:temp-scale}), is $3.82 \pm 0.12$.

Gray shaded regions show the residual scatter about the mean relation.  There are a small number of outliers, particularly toward low values of $T_X$ given $\lamRM$ or, equivalently, a larger $\lamRM$ than expected given their $T_X$.  Such systems are likely to have a boosted richness due to lower-mass halos along the line of sight that boost $\lamRM$ more than $T_X$ \citep{Rozo:2014, Ge:2018}.  

We test the shape of $\Pr({\rm k} T_X \,|\, \lambda, z)$ by examining the normalized residuals of the data about the best-fit mean scaling, 
\begin{equation}\label{eq:residual}
\delta_{T|\lambda\,,\,i} = \frac{\ln \left( E^{-2/3}(z_i) \, {\rm k}T_{i} \right) - \alpha_{T|\lambda} \ln(\lambda_{i}) - \pi_{T|\lambda}}{ (\varTxlambda + \sigma_{{\rm err}\,,\,i}^2 )^{1/2} }, 
\end{equation}
using posterior maximum likelihood estimates of the parameters $\pi_{T|\lambda}$, $\alpha_{T|\lambda}$ and $\varTxlambda$, and the index $i$ corresponds to the $i^{\rm th}$ cluster. The quadrature inclusion of $\sigma_{{\rm err}, i}^2$, the square of the  measurement uncertainty in $\ln T_i$, is appropriate if measurement errors are both accurately estimated and also uncorrelated with the underlying astrophysical processes responsible for the residual scatter.

\begin{figure}
    \centering
    \includegraphics[width=0.23\textwidth]{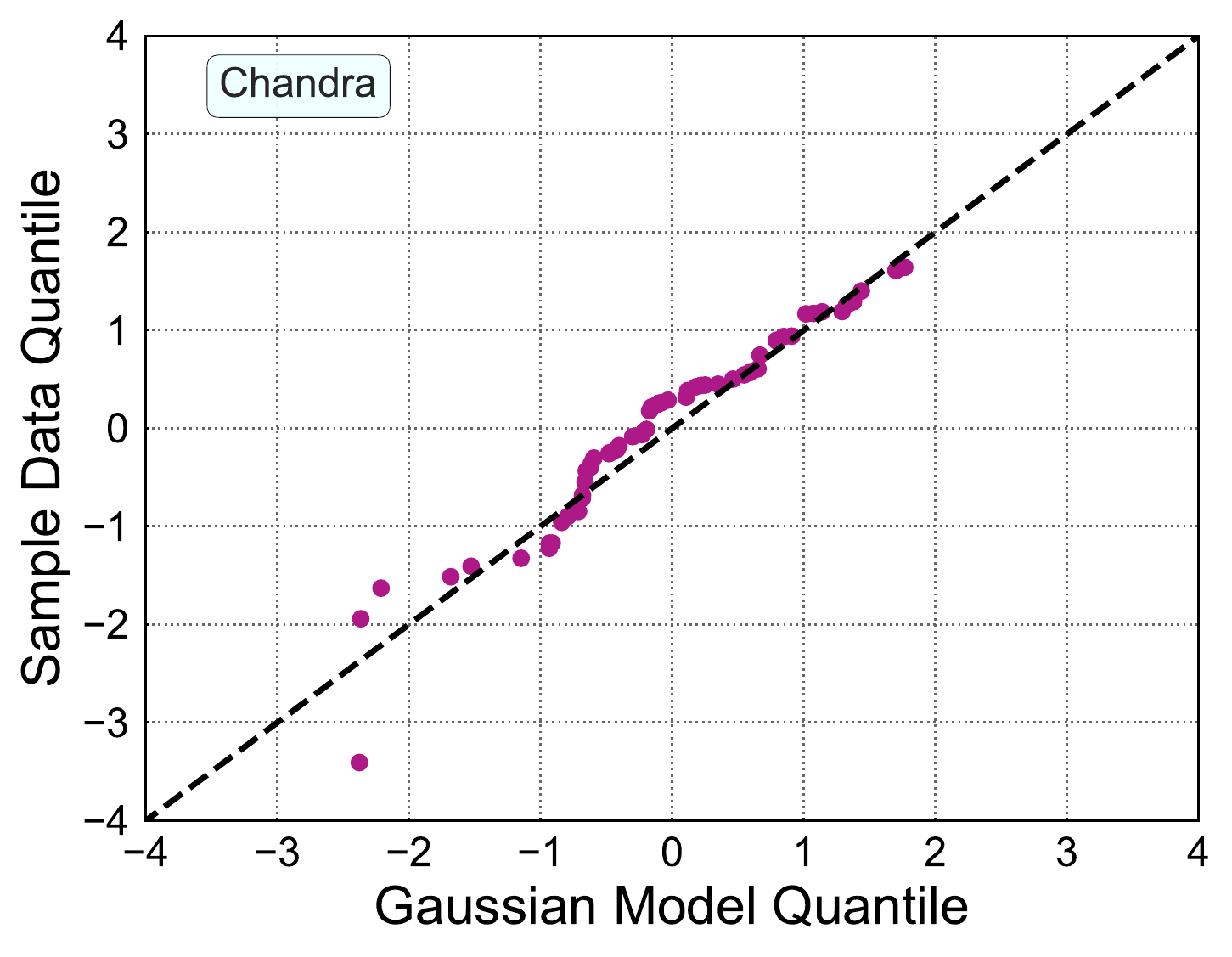}
    \includegraphics[width=0.23\textwidth]{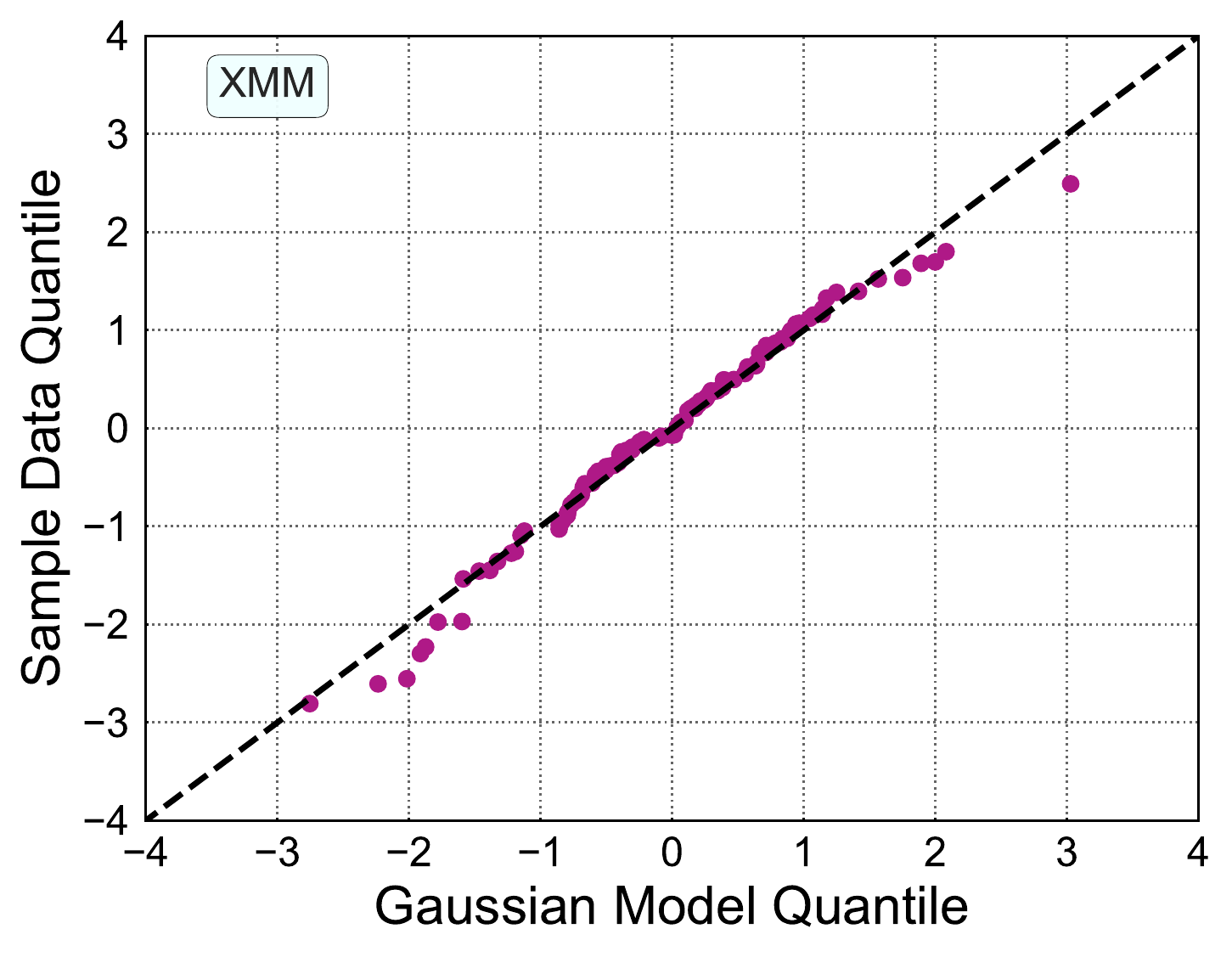} 
    \includegraphics[width=0.35\textwidth]{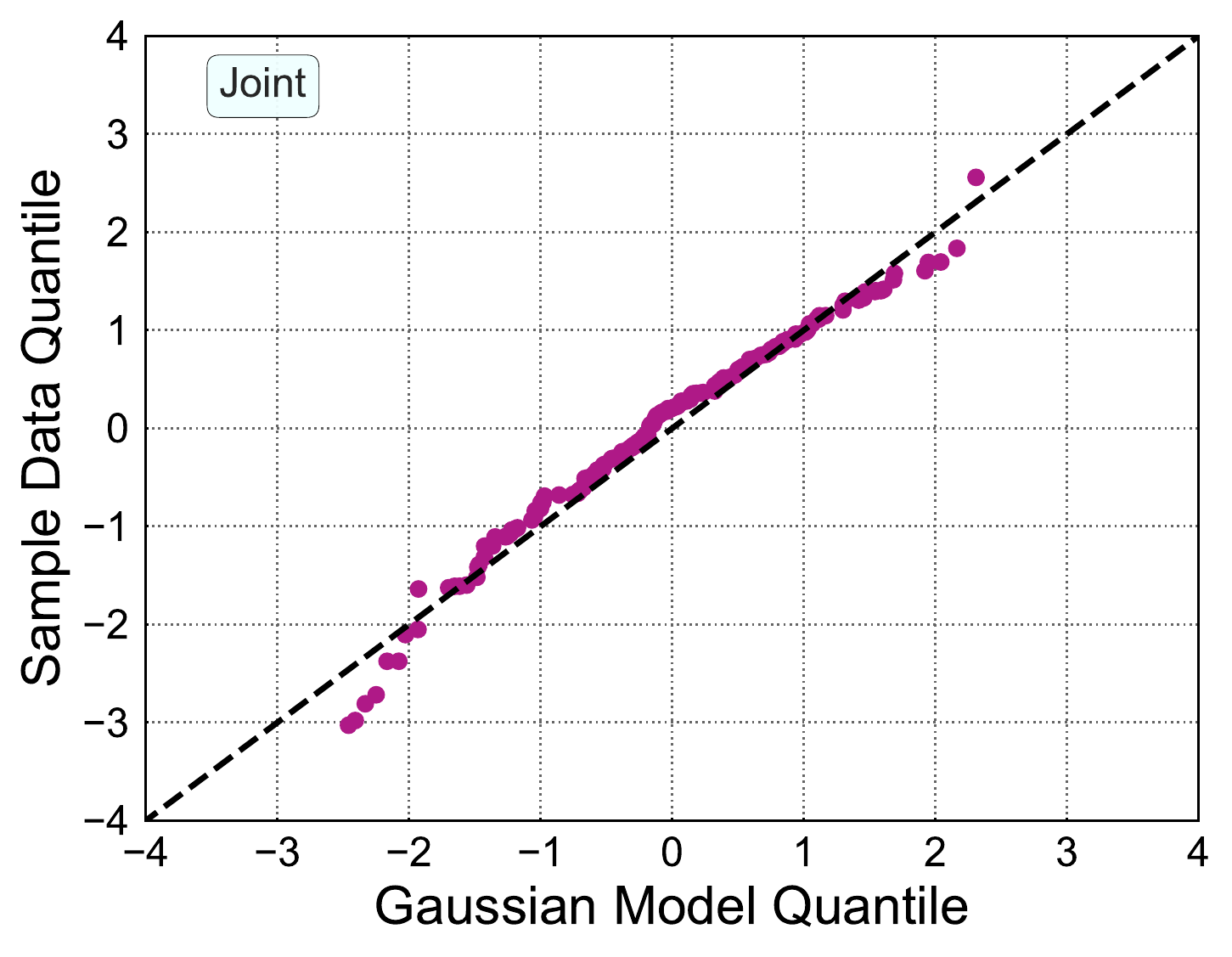}
    \caption{Normalized, ranked residuals, equation~(\ref{eq:residual}), of the \chandra\ (top, left), \xmm\ (top, right), and joint (bottom) samples follow closely a log-normal form, as indicated by the close proximity of the points to the dashed line of equality.}
 \label{fig:qq-plot}
\end{figure}

Figure~\ref{fig:qq-plot} shows quantile-quantile (Q-Q) plots of the residuals in both samples. The Q-Q plot compares the quantiles of the rank-ordered residuals, expressed in units of the measured standard deviation, equation~(\ref{eq:residual}), to those expected under the assumed Gaussian model.  The Q-Q form of both samples support the log-normal likelihood, equation~(\ref{eq:ProbTxlambda}), as shown by the proximity of the measured quantiles to the dashed line of unity.  There is a very slight skew in the distribution, with more weight to the low-temperature side as expected from projection effects \citep{Cohn:2007}. In a previous work, \citet{Mantz:2008} employed the Q-Q plot and illustrated the log-normality of their cluster sample (see their Fig. 4).

Within the statistical uncertainties, the slope of temperature--richness scaling is consistent with a simple self-similar expectation of $2/3$, the result obtained if the star formation efficiency is constant, so that $\lambda \propto M_{\rm star} \propto M$, and the temperature scales as $T \propto [E(z) M]^{2/3}$ from virial equilibrium \citep{Kaiser:1991, BryanNorman:1998}. However, both dynamical and weak lensing analysis of the same SDSS sample produce mean behavior $M \propto \lamRM^{1.3 \pm 0.1}$, which shows deviation from the self-similar expectation \citep{Simet:2017,Farahi:2016}, and the Weighing the Giants (WtG) analysis yielding $T_X \propto M^{0.62 \pm 0.04}$.  

If $\rlambdat$ is close to zero and there is no scatter about the mean relation, then it is expected that temperature scales with mass with slope $\alpha_{ T \, | \, \lambda} = 0.81 \pm 0.10$.  There is moderate tension with our result of $0.62 \pm 0.04$ for the joint sample, which we suspect reflects the lack of low luminosity and temperature systems having low optical richness in the X-ray archives.

\subsection{Mass scatter conditioned on optical richness} \label{sec:halo-prop}

\begin{table}
\centering
\caption{External constraints required for the richness-conditioned mass variance, equation~(\ref{eq:varmulambda}).  Uncertainties are $68\%$ confidence intervals.}
\label{tab:externalParams}
\begin{tabular}{|l|c|c|}
\hline
Parameter & Value & Reference \\ \hline
$\sigma_{\ln T \, |\, M }$ & $0.16 \pm 0.02$ & \citet{Mantz:2016WtG-V}  \\ 
$\alpha_{T \, |\, M }$ & $0.62 \pm 0.04$ & \citet{Mantz:2016WtG-V}  \\ 
$\sigmamuTx $ & $0.26 \pm 0.04$   & (inferred from above)   \\ 
$\rlambdat$    & $-0.25^{+0.24}_{-0.22}$ & Farahi et al. (in preparation)  \\ \hline
\end{tabular}
\end{table}

To derive mass variance with the multi-property population framework, information on additional scaling parameters is required, as reflected in equation~(\ref{eq:varmulambda}).  We employ recent derivations of scaling behavior from the WtG program \citep{Mantz:2015:wtgI}, particularly the slope and residual scatter of the $T_X$--$M_{\rm wl}$ relation derived by \citet{Mantz:2016WtG-V}.  In the WtG analysis, X-ray properties are regressed against lensing mass estimates for a sample of 40 clusters. Their posterior estimates of the scatter in temperature, $\sigma_{\ln T \, | \,\mu} = 0.16 \pm 0.02$, and slope, $\alpha_{T \, |\, \mu} = 0.62 \pm 0.04$, imply a scatter in mass at fixed temperature of $0.26 \pm  0.04$. These values are listed in Table~{\ref{tab:externalParams}}, and we assume that the uncertainties are Gaussian distributed.  

For the  correlation coefficient between temperature and richness at fixed halo mass, we use constraints derived from nearby LoCuSS clusters by Farahi et al. (in preparation).  The X-ray bright LoCuSS sample \citep{Mulroy:2019} contains 41 clusters in the redshift range $0.15 < z < 0.3$, with 33 overlapping the SDSS sample region. The \RM\ richness estimates for those systems range from 27 to 181, with the median value near 100.  Using the same model framework as this paper, modified to include the original X-ray selection criteria, Farahi et al. (in preparation) derive the first empirical constraint on the correlation coefficient, $\rlambdat = -0.25^{+0.24}_{-0.22}$.  

\begin{figure}
    \centering
    \includegraphics[width=0.48\textwidth]{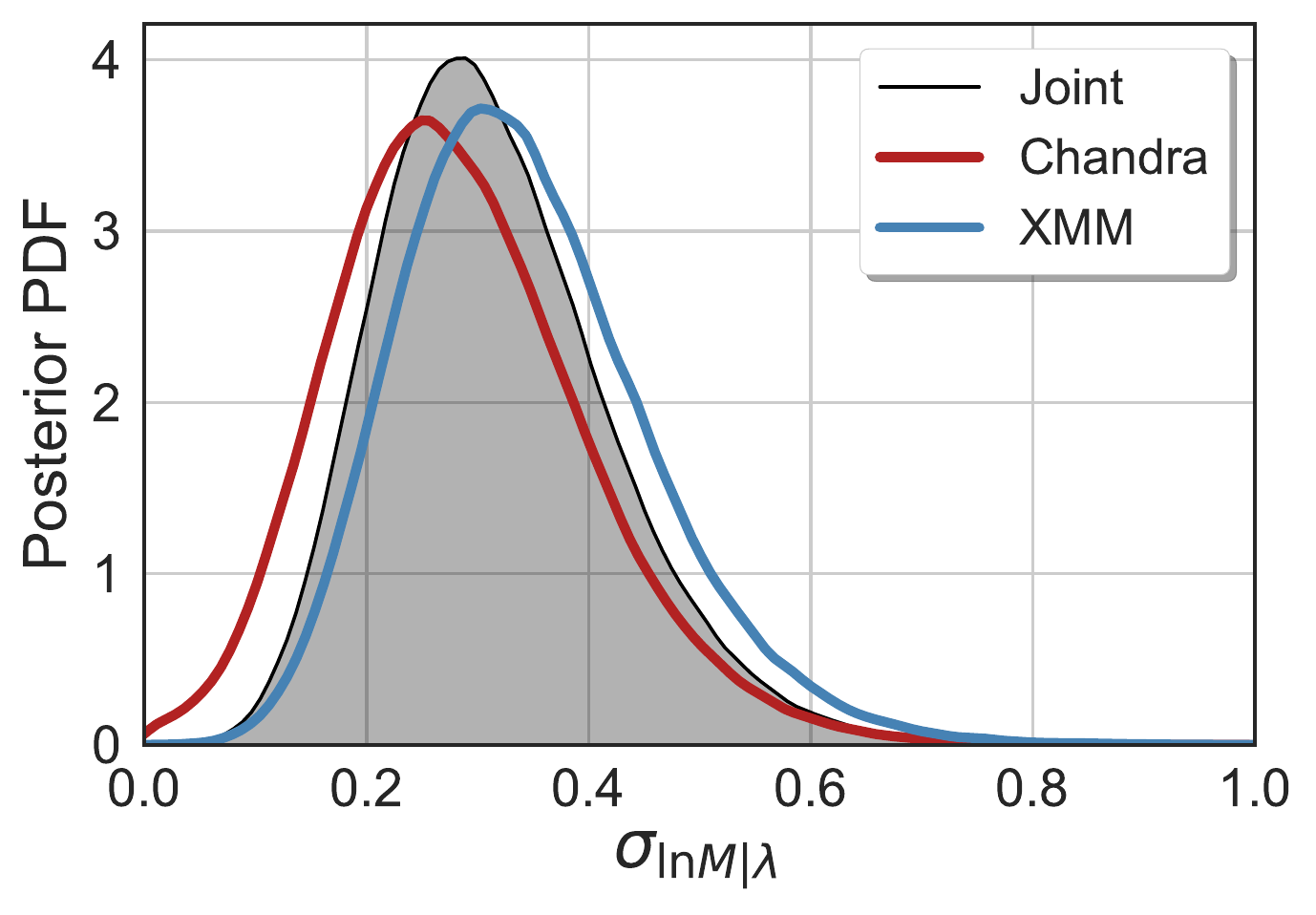}
    \caption{Marginalized posterior likelihood of the halo mass scatter at fixed richness derived from the supplementary X-ray samples given in the legend. }
    \label{fig:post-sigma}
\end{figure}

With these additional elements, we can now derive an estimate of the richness-conditioned mass scatter, equation~(\ref{eq:varmulambda}), resulting in $\sigmamulambda = 0.30 \pm 0.10$. To get this result, we employ the values derived from the joint sample temperature variance conditioned on richness.  
Since the correlation coefficient is broad and slightly asymmetric (see Fig. \ref{fig:prior-r}), we use Monte Carlo sampling of the terms on the right-hand side of equation~(\ref{eq:varmulambda}), discarding any combinations that produce unphysical results (negative values inside the square root).  The resultant posterior distributions are shown in Fig. \ref{fig:post-sigma} for the \chandra, \xmm, and joint analysis with values of $\varTxlambda$ from Table~\ref{tab:TxlambdaParams}.  

For the joint sample analysis, the median value of the posterior mass scatter is
\begin{equation} \label{eq:sigmaMfinal}
\sigma_{\ln M| \lambda} \ = \ 0.30\,\pm 0.04\, _{({\rm stat})} \pm 0.09\, _{({\rm sys})}  ,
\end{equation}
where the quoted uncertainties are $68\%$ confidence level.  The statistical error derives from the $T_X$--$\lamRM$ residual variance uncertainty while the systematic uncertainty is derived from the quoted errors of the required external parameters. The overall 95\% confidence region is broad, spanning $0.14$ to $0.55$.  


\section{Discussion}\label{sec:discussion}

Here we review our treatment of systematic uncertainties, including priors, before turning to a comparison with existing estimates of the mass scatter using optical proxies in both observed cluster samples and simulated halo ensembles.

\subsection{Systematic and prior uncertainties}

The richness-conditioned mass variance is inferred from the observed temperature variance via equation~(\ref{eq:varmulambda}).  Uncertainties in the additional elements (see Table~\ref{tab:externalParams}) propagates to broaden the uncertainty in $\sigmalnMlambda$. 

\begin{figure}
    \centering
    \includegraphics[width=0.48\textwidth]{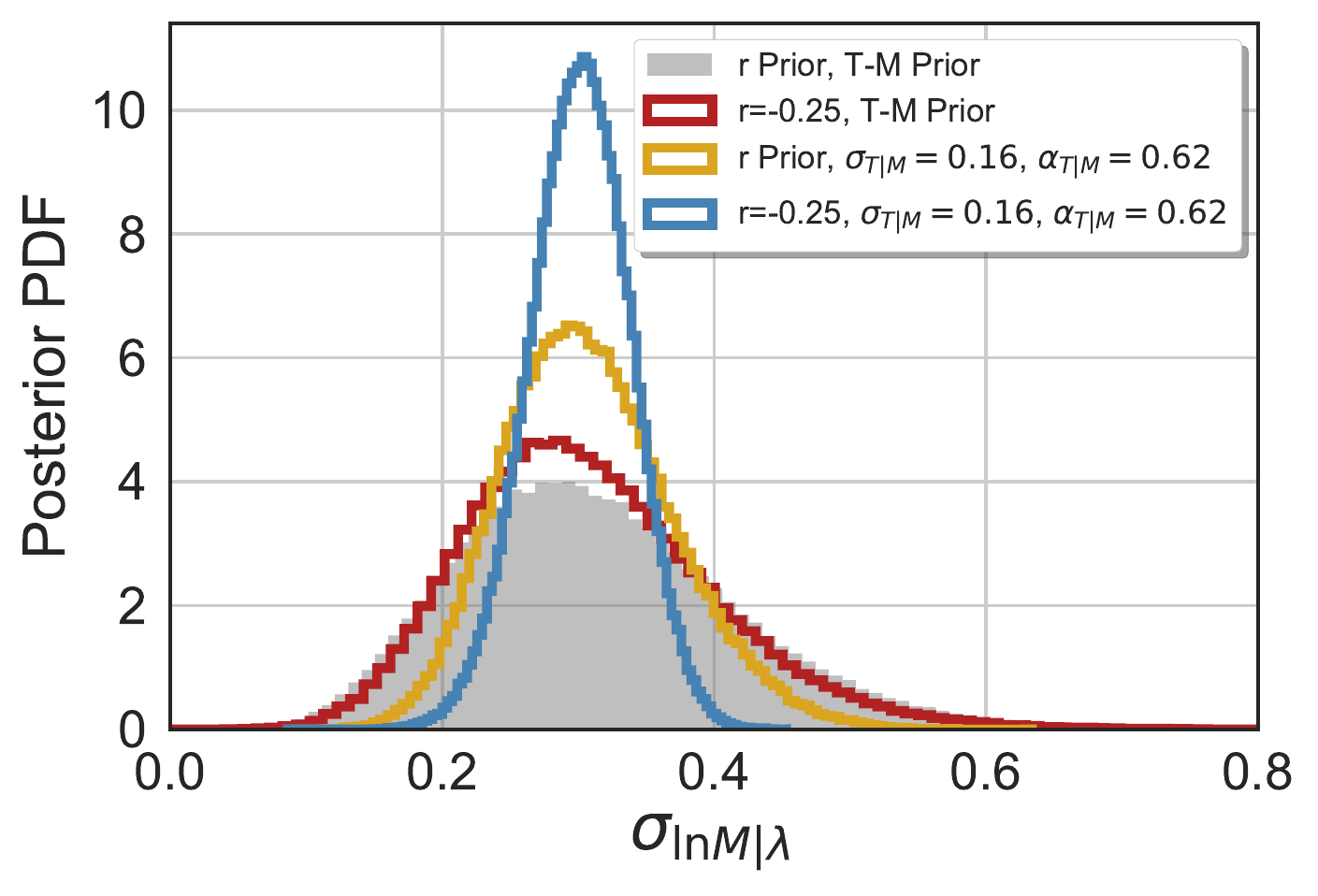}
    \caption{Illustration of the impact of uncertainties in the external elements, Table~\ref{tab:externalParams}.  The grey shaded region shows the joint analysis result fully marginalized over external uncertainties.  Red and yellow lines assume, respectively, that $\rlambdat$ or the $T_X$--$M$ scaling parameters are exactly known.  The $T_X$--$M$ parameters dominate the error budget. The blue line shows the case of all external quantities known. The width of the blue is our statistical uncertainties.}
    \label{fig:systematics}
\end{figure}

Figure \ref{fig:systematics} explores the contribution of each element's uncertainty by systematically setting the error in specific terms to zero -- i.e., fixing the value of each element.  The yellow curve fixes the $T_{X}$--$M_{\rm wl}$ relation parameters, both slope and scatter, while the red curve fixes the correlation coefficient, $\rlambdat$.  Finally, the blue curve shows the impact of having perfect knowledge of all above parameters.  

The temperature--mass relation uncertainties make the largest contribution to the uncertainty in mass variance. We note that the contribution of the temperature--mass uncertainty in the slope is negligible and it is dominated by the uncertainty in the scatter. The width of the blue curve, the statistical uncertainty, is 0.04.

A potential source of systematic uncertainty in our analysis is the lack of a selection model for the archival X-ray analysis.  The modest sampling of $\lamRM < 100$ (see Fig.~\ref{fig:e-selection-function}) leaves open the possibility that only the X-ray brightest subset of clusters are being measured at low optical richness.  Such sample truncation in flux would be likely to influence temperature selection, eliminating low-$T_X$ clusters and thereby distorting estimates of both the slope and variance at low richness values.   We defer a detailed study of these effects to future work.

We show in Appendix~\ref{app:running} that the current data do not display strong dependence on an imposed lower richness limit, and a conservative view is to interpret the mass scatter constraints we report as appropriate to $\lamRM = 100$, since the supplementary samples are, cumulatively, 50\% complete above this richness. We note that Poisson scatter sets the minimum expected richness variance at fixed mass. This implies minimum mass scatter of $0.1 \, \alpha_{\lnM \, | \, \lambda} \simeq 0.13$, using a mass--richness scaling, $\langle \ln M | \lambda \rangle \propto \lambda^{1.3}$ \citep{Simet:2017,Farahi:2016}. This value lies just outside the $2 \sigma$ low tail of our posterior joint-sample constraints.

\subsection{Comparing to previous studies}

Using a smaller set of \RM\ clusters identified in science verification (DES-SV) data, \citet{Rykoff:2016} estimated the slope and the scatter of $T_X$--$\lamRM$ scaling relation to be $0.60 \pm 0.09$ and $0.28^{+0.07}_{-0.05}$, respectively.  With a larger sample size, we find values in agreement with those of \citet{Rykoff:2016}, but with smaller parameter uncertainties.  

Our measurements can also be compared to those of \citet{Rozo:2014}.  That work set is similar in spirit to this study but differs in some key details. The correlation coefficient, $\rlambdat$, was set to zero, and the analysis did not propagate uncertainties in the $T_X$--$M$ relation. Employing sub-samples of X-ray selected clusters from the literature, including the XCS \citep{Mehrtens:2012}, MCXC \citep{Piffaretti:2011}, ACCEPT \citep{Cavagnolo:2009}, and \citet{Mantz:2010} cluster samples, they estimate $\sigmalnMlambda = 0.26 \pm 0.03$, with the quoted uncertainty being entirely statistical.  Our central value is consistent with theirs, but a key step of our analysis is to more carefully revise uncertainties by incorporating a coherent multi-property model.

In a separate work, \citet{Rozo:2015} directly estimated the scatter in richness at fixed SZ-mass by comparing the \RM\ catalog to the \emph{Planck} SZ-selected cluster catalog \citep{Planck:2014}. They estimate $\sigmalnMlambda = 0.277 \pm 0.026$, with the reported uncertainties again being purely statistical.  We note that the SZ-masses are inferred from $Y_{\rm SZ}$--$M$ relation, so covariance between $\lamRM$ and $Y_{\rm SZ}$ needs to be taken into account in this analysis.

In an independent analysis using abundance and stacked weak lensing profiles for roughly 8,000 SDSS \RM\ clusters with richness, $20 \le \lamRM  \le 100$, and redshift, $0.1 < z < 0.33$, \citet{Murata:2017} derive $\sigmalnMlambda \sim 0.46 \pm 0.05$ at a pivot mass scale of $3 \times 10^{14} \hinv \msun$, equivalent to a richness of $24$, from their mean scaling relation.   In their analysis, the scatter is allowed to run with mass, and they find that $\sigmalnMlambda \propto M^{-0.17 \pm 0.03}$. Evaluating their result at a richness of 70, or a mass scale roughly a factor of 4 larger, leads to a mass scatter of $0.36$, consistent with our findings.

This work is concerned about mass scatter conditioned on optical-richness. To estimate the richness scatter at fixed halo mass, \citet{Saro:2015} modeled the total richness variance conditioned on halo mass with a Poisson term and a log-normal scatter term. If this additional Poisson contribution, at pivot richness of our joint sample $\lambda_{\rm piv} \sim 72$, is subtracted from the total variance, the richness variance conditioned on the halo mass yields
\begin{equation} \label{eq:scatter-final}
     {\rm Var}(\ln \lambda | M) \equiv \frac{ {\rm Var}(\mu | \ln \lambda) }{ \alpha^2_{\mu | \ln \lambda} } = \exp(- \langle \ln \lambda | M \rangle) + \sigma^2_{\ln \lambda | M}, 
\end{equation}
where ${\rm Var}(\mu | \ln \lambda) = 0.093^{+0.082}_{-0.047}$ is the halo mass variance conditioned on optical-richness, and $\alpha^2_{\mu | \ln \lambda} = 1.356 \pm 0.052$ is the slope of $M$--$\lamRM$ relation \citep{DESY1-WL:2018}. Plugging these numbers into Eq. \ref{eq:scatter-final}, we infer
\begin{equation}
   \sigma_{\ln \lambda | M} = 0.20^{+0.10}_{-0.08}.
\end{equation}
This result is consistent with what is previously found employing \RM\ clusters from SDSS survey \citep[][]{Saro:2015,Simet:2017,SDSS:2018cosmology}.

\subsection{Redshift dependence}

We find no evidence of the redshift evolution for the slope and the scatter of the $T_X$--$\lamRM$ relation. We split the \chandra, the \xmm, and the joint samples in half at $z=0.4$. The $T_X$--$\lamRM$ relation results are presented in Table \ref{tab:TxlambdaParamsZSplit}.  

\citet{Farahi:2017} studied the redshift evolution of integrated stellar mass -- halo mass scaling relation employing the hydrodynamical simulations.  They find a mild redshift evolution for both slope and the scatter of this relation. The statistical uncertainties of our sample are larger than the magnitude of the redshift evolution they noticed. Therefore, we cannot rule out or confirm such a small evolution using the current sample.

\begin{table}
\begin{center}
\caption{Best-fit parameters for the $T_X$--$\lamRM$ relation. Samples are splitted into two non-overlapping subsets with $z>4$ and $z \le 0.4$. The notation is similar to Table \ref{tab:TxlambdaParams}.} 
\label{tab:TxlambdaParamsZSplit}
\begin{tabular}{|l|c|c|c|}
\hline
  & Slope & Residual scatter \\ 
Sample  & $\alpha_{T \, | \, \lambda }$ & $\sigma_{\ln T | \lambda}$ \\ \hline
\chandra\ ($z \leq 0.4$) & $0.60 \pm 0.13$      & $0.30 \pm 0.06$  \\ 
\xmm\ ($z\leq 0.4$)      & $0.56 \pm 0.07$      & $0.27 \pm 0.04$   \\
Joint ($z\leq 0.4$)      & $0.59 \pm 0.05$      & $0.27 \pm 0.03$   \\ \hline
\chandra\ ($z>0.4$)      & $0.51 \pm 0.15$      & $0.25 \pm 0.05$  \\ 
\xmm\ ($z>0.4$)          & $0.65 \pm 0.08$      & $0.32 \pm 0.04$   \\ 
Joint ($z>0.4$)          & $0.65 \pm 0.06$      & $0.29 \pm 0.03$   \\ \hline
\end{tabular}
\end{center}
\end{table}

\subsection{Application to DES cluster cosmology}

The mass variance constraints we derive can inform priors for cluster cosmology studies.  For the DES survey, 
the model that links observed cluster richness with halo mass \citep{SDSS:2018cosmology} is more complex than the log-normal population model we apply here. In particular, \citet{Costanzi:2018projection} develop an explicit model of projection that is a component of a hierarchical Bayes framework for $\Pr (\lamRM | M, z )$.  The base of that framework is an intrinsic halo population variance frames as a Poisson distribution convolved with a Gaussian of width, $\sigma_{\rm intr}$.  

However, \citet{SDSS:2018cosmology} find that a log-normal model for the intrinsic halo population gives cosmological constraints consistent with the Poisson plus Gaussian model, and posterior estimates of $\Pr (M | \lamRM, z )$ are found to be nearly log-normal.  More work is needed to fully incorporate constraints of the type derived in this study into cosmological analysis pipelines. 


\section{Conclusion} \label{sec:conclusion}

We use archival X-ray observations of 168 \RM\ clusters identified in DES-Y1 imaging to place limits on the mass variance at fixed galaxy richness, a critical component of cluster cosmology analysis.  The X-ray observables, $T_X$ and $L_X$, of galaxy clusters at redshifts $0.2 < z < 0.7$ falling within archival \chandra\ or \xmm\ archival data are extracted via MATCha and XAPA processing pipelines, respectively.  We determine parameters of a power-law $T_X$--$\lamRM$ relation, particularly the residual scatter in the log of temperature conditioned on richness, and infer the halo mass scatter at fixed optical richness using a log-normal multi-property population model.  

Given the modest sample size and the lack of a detailed X-ray selection model, we do not attempt to add scaling of the mass variance with cluster richness or redshift. The median redshift of both samples is $0.41$ while the median richness is 76 for \chandra\ and 47 for the larger \xmm\ sample.  We infer residual scatter in temperature at fixed richness, $\sigmaTxlambda = 0.26 \pm 0.03$ (\chandra) and $0.29 \pm 0.03$ (\xmm).  The moderately larger variance in the lower-richness \xmm\ sample may be providing a hint of mass dependence. Larger samples and a model for archival X-ray selection are required to address this issue.

Constraining the mass scatter requires additional information: the slope and variance of the $T_X$--$M$ relation as well as the correlation between $\lamRM$ and $T_X$ at fixed halo mass. Incorporating values from the Weighing the Giants and LoCuSS samples, respectively, and using the richness-conditioned temperature variance from the combined sample, we derive the mass scatter parameter, $\sigma_{\ln M| \lambda} \ = \ 0.30\,\pm 0.04\, _{({\rm stat})} \pm 0.09\, _{({\rm sys})}$.  

The contribution of the external parameter uncertainties in these systematics to the overall uncertainty budget is considerable. Therefore, as we make progress to better understand the scaling relations of multi-wavelength observables, it is necessary to pay attention to the off-diagonal elements of the mass-conditioned property covariance matrix.  \citet{Mantz:2016-relaxedIII} pioneered empirical estimates of the full covariance matrix for X-ray observables and Farahi et al. (in preparation) take the lead in combining optical, X-ray and SZ observables in the LoCuSS sample. Improved understanding of the broad property covariance matrix behavior will allow us to improve the mass variance constraints from studies such as this.

\section*{Acknowledgments}

A. Farahi is supported by a McWilliams Postdoctoral Fellowship. T. Jeltema is pleased to acknowledge funding support from DE-SC0010107 and DE-SC0013541.

Funding for the DES Projects has been provided by the U.S. Department of Energy, the U.S. National Science Foundation, the Ministry of Science and Education of Spain, the Science and Technology Facilities Council of the United Kingdom, the Higher Education Funding Council for England, the National Center for Supercomputing Applications at the University of Illinois at Urbana-Champaign, the Kavli Institute of Cosmological Physics at the University of Chicago, the Center for Cosmology and Astro-Particle Physics at the Ohio State University, the Mitchell Institute for Fundamental Physics and Astronomy at Texas A\&M University, Financiadora de Estudos e Projetos,  Funda{\c c}{\~a}o Carlos Chagas Filho de Amparo {\`a} Pesquisa do Estado do Rio de Janeiro, Conselho Nacional de Desenvolvimento Cient{\'i}fico e Tecnol{\'o}gico and the Minist{\'e}rio da Ci{\^e}ncia, Tecnologia e Inova{\c c}{\~a}o, the Deutsche Forschungsgemeinschaft and the Collaborating Institutions in the Dark Energy Survey. 

The Collaborating Institutions are Argonne National Laboratory, the University of California at Santa Cruz, the University of Cambridge, Centro de Investigaciones Energ{\'e}ticas, Medioambientales y Tecnol{\'o}gicas-Madrid, the University of Chicago, University College London, the DES-Brazil Consortium, the University of Edinburgh, the Eidgen{\"o}ssische Technische Hochschule (ETH) Z{\"u}rich, Fermi National Accelerator Laboratory, the University of Illinois at Urbana-Champaign, the Institut de Ci{\`e}ncies de l'Espai (IEEC/CSIC), the Institut de F{\'i}sica d'Altes Energies, Lawrence Berkeley National Laboratory, the Ludwig-Maximilians Universit{\"a}t M{\"u}nchen and the associated Excellence Cluster Universe, the University of Michigan, the National Optical Astronomy Observatory, the University of Nottingham, The Ohio State University, the University of Pennsylvania, the University of Portsmouth,  SLAC National Accelerator Laboratory, Stanford University, the University of Sussex, Texas A\&M University, and the OzDES Membership Consortium.

Based in part on observations at Cerro Tololo Inter-American Observatory, National Optical Astronomy Observatory, which is operated by the Association of Universities for Research in Astronomy (AURA) under a cooperative agreement with the National Science Foundation.

The DES data management system is supported by the National Science Foundation under Grant Numbers AST-1138766 and AST-1536171. The DES participants from Spanish institutions are partially supported by MINECO under grants AYA2015-71825, ESP2015-66861, FPA2015-68048, SEV-2016-0588, SEV-2016-0597, and MDM-2015-0509, some of which include ERDF funds from the European Union. IFAE is partially funded by the CERCA program of the Generalitat de Catalunya. Research leading to these results has received funding from the European Research Council under the European Union's Seventh Framework Program (FP7/2007-2013) including ERC grant agreements 240672, 291329, and 306478. We  acknowledge support from the Brazilian Instituto Nacional de Ci\^enciae Tecnologia (INCT) e-Universe (CNPq grant 465376/2014-2).

This manuscript has been authored by Fermi Research Alliance, LLC under Contract No. DE-AC02-07CH11359 with the U.S. Department of Energy, Office of Science, Office of High Energy Physics. The United States Government retains and the publisher, by accepting the article for publication, acknowledges that the United States Government retains a non-exclusive, paid-up, irrevocable, world-wide license to publish or reproduce the published form of this manuscript, or allow others to do so, for United States Government purposes.

\section*{Affiliations}
{\small
$^{1}$  McWilliams Center for Cosmology, Department of Physics, Carnegie Mellon University, Pittsburgh, Pennsylvania 15312, USA\\
$^{2}$ Department of Physics, University of Michigan, Ann Arbor, MI 48109, USA\\
$^{3}$ Department of Physics, Yale University, New Haven, CT 06511, USA\\
$^{4}$ Department of Astronomy, University of Michigan, Ann Arbor, MI 48109, USA\\
$^{5}$ Santa Cruz Institute for Particle Physics, Santa Cruz, CA 95064, USA\\
$^{6}$ Department of Physics and Astronomy, Pevensey Building, University of Sussex, Brighton, BN1 9QH, UK\\
$^{7}$ Astrophysics \& Cosmology Research Unit, School of Mathematics, Statistics \& Computer Science, University of KwaZulu-Natal, Westville Campus, Durban 4041, South Africa\\
$^{8}$ Department of Physics, University of Arizona, Tucson, AZ 85721, USA\\
$^{9}$ Kavli Institute for Particle Astrophysics \& Cosmology, P. O. Box 2450, Stanford University, Stanford, CA 94305, USA\\
$^{10}$ SLAC National Accelerator Laboratory, Menlo Park, CA 94025, USA\\
$^{11}$ Astrophysics Research Institute, Liverpool John Moores University, IC2, Liverpool Science Park, 146 Brownlow Hill, Liverpool, L3 5RF, UK\\
$^{12}$ Universit\"ats-Sternwarte, Fakult\"at f\"ur Physik, Ludwig-Maximilians Universit\"at M\"unchen, Scheinerstr. 1, 81679 M\"unchen, Germany\\
$^{13}$ CENTRA, Instituto Superior T\'ecnico, Universidade de Lisboa, Av. Rovisco Pais 1, 1049-001 Lisboa, Portugal\\
$^{14}$ Institute for Astronomy, University of Edinburgh, Edinburgh EH9 3HJ, UK\\
$^{15}$ Perimeter Institute for Theoretical Physics, 31 Caroline St. North, Waterloo, ON N2L 2Y5, Canada\\
$^{16}$ BIPAC, Department of Physics, University of Oxford, Denys Wilkinson Building, 1 Keble Road, Oxford OX1 3RH, UK\\
$^{17}$ Sub-department of Astrophysics, Department of Physics, University of Oxford, Denys Wilkinson Building, Keble Road, Oxford OX1 3RH, UK and Department of Physics, Lancaster University, Lancaster LA1 4 YB, UK\\
$^{18}$ Instituto de Astrofisica e Ciencias do Espaco, Universidade do Porto, CAUP, Rua das Estrelas, P-4150-762 Porto, Portugal \\
$^{19}$ Fermi National Accelerator Laboratory, P. O. Box 500, Batavia, IL 60510, USA\\
$^{20}$ Institute of Cosmology and Gravitation, University of Portsmouth, Portsmouth, PO1 3FX, UK\\
$^{21}$ Department of Physics \& Astronomy, University College London, Gower Street, London, WC1E 6BT, UK\\
$^{22}$ Centro de Investigaciones Energ\'eticas, Medioambientales y Tecnol\'ogicas (CIEMAT), Madrid, Spain\\
$^{23}$ Laborat\'orio Interinstitucional de e-Astronomia - LIneA, Rua Gal. Jos\'e Cristino 77, Rio de Janeiro, RJ - 20921-400, Brazil\\
$^{24}$ Department of Astronomy, University of Illinois at Urbana-Champaign, 1002 W. Green Street, Urbana, IL 61801, USA\\
$^{25}$ National Center for Supercomputing Applications, 1205 West Clark St., Urbana, IL 61801, USA\\
$^{26}$ Institut de F\'{\i}sica d'Altes Energies (IFAE), The Barcelona Institute of Science and Technology, Campus UAB, 08193 Bellaterra (Barcelona) Spain\\
$^{27}$ Institut d'Estudis Espacials de Catalunya (IEEC), 08034 Barcelona, Spain\\
$^{28}$ Institute of Space Sciences (ICE, CSIC),  Campus UAB, Carrer de Can Magrans, s/n,  08193 Barcelona, Spain\\
$^{29}$ Observat\'orio Nacional, Rua Gal. Jos\'e Cristino 77, Rio de Janeiro, RJ - 20921-400, Brazil\\
$^{30}$ Department of Physics, IIT Hyderabad, Kandi, Telangana 502285, India\\
$^{31}$ Excellence Cluster Origins, Boltzmannstr.\ 2, 85748 Garching, Germany\\
$^{32}$ Faculty of Physics, Ludwig-Maximilians-Universit\"at, Scheinerstr. 1, 81679 Munich, Germany\\
$^{33}$ Kavli Institute for Cosmological Physics, University of Chicago, Chicago, IL 60637, USA\\
$^{34}$ Instituto de Fisica Teorica UAM/CSIC, Universidad Autonoma de Madrid, 28049 Madrid, Spain\\
$^{35}$ Department of Physics, Stanford University, 382 Via Pueblo Mall, Stanford, CA 94305, USA\\
$^{36}$ Center for Cosmology and Astro-Particle Physics, The Ohio State University, Columbus, OH 43210, USA\\
$^{37}$ Department of Physics, The Ohio State University, Columbus, OH 43210, USA\\
$^{38}$ Harvard-Smithsonian Center for Astrophysics, Cambridge, MA 02138, USA\\
$^{39}$ Department of Astronomy/Steward Observatory, University of Arizona, 933 North Cherry Avenue, Tucson, AZ 85721-0065, USA\\
$^{40}$ Australian Astronomical Optics, Macquarie University, North Ryde, NSW 2113, Australia\\
$^{41}$ Departamento de F\'isica Matem\'atica, Instituto de F\'isica, Universidade de S\~ao Paulo, CP 66318, S\~ao Paulo, SP, 05314-970, Brazil\\
$^{42}$ George P. and Cynthia Woods Mitchell Institute for Fundamental Physics and Astronomy, and Department of Physics and Astronomy, Texas A\&M University, College Station, TX 77843,  USA\\
$^{43}$ Department of Astrophysical Sciences, Princeton University, Peyton Hall, Princeton, NJ 08544, USA\\
$^{44}$ Instituci\'o Catalana de Recerca i Estudis Avan\c{c}ats, E-08010 Barcelona, Spain\\
$^{45}$ School of Physics and Astronomy, University of Southampton,  Southampton, SO17 1BJ, UK\\
$^{46}$ Instituto de F\'isica Gleb Wataghin, Universidade Estadual de Campinas, 13083-859, Campinas, SP, Brazil\\
$^{47}$ Computer Science and Mathematics Division, Oak Ridge National Laboratory, Oak Ridge, TN 37831\\
$^{48}$ Argonne National Laboratory, 9700 South Cass Avenue, Lemont, IL 60439, USA\\
$^{49}$ Cerro Tololo Inter-American Observatory, National Optical Astronomy Observatory, Casilla 603, La Serena, Chile\\
$^{50}$ Max Planck Institute for Extraterrestrial Physics, Giessenbachstrasse, 85748 Garching, Germany\\
}

\bibliographystyle{mn2e_adsurl}
\bibliography{apj-jour,astroref}

\appendix

\section{Cluster Catalogs} \label{app:cluster-cat}

In Table \ref{tab:chandra-tab} and \ref{tab:XMM-table}, we provide the optical and the X-ray properties of \chandra\ and \xmm\ clusters, i.e. the data vector, employed in this work. The \texttt{MEM\_MATCH\_ID} is the \RM\ Cluster Identification Number that shall be used to match the X-ray clusters to the original \RM\ clusters \citep{DESY1-WL:2018}. The full original \RM\ DES Y1A1 catalogs will be available at \url{http://risa.stanford.edu/redmapper/} in FITS format. \texttt{LAMBDA\_CHISQ} and \texttt{LAMBDA\_CHISQ (X-ray peak)} are the original \RM\ optical richness and the new richness assigned to each cluster at the location of the X-ray emission peak, respectively. \texttt{XCS\_NAME} in Table \ref{tab:XMM-table} is the unique source identifier which could be used to match with the XCS source catalog (Giles et al. in preparation). The full X-ray catalogs will be available from the online journal in machine-readable formats.

\begin{table*}
\caption{\chandra\ Clusters.}\label{tab:chandra-tab}
\begin{tabular}{|l|l|l|l|l|l|}
\hline
MEM\_MATCH\_ID & $z_{\lambda}$ & LAMBDA\_CHISQ & LAMBDA\_CHISQ (X-ray peak) & ${\rm k}T_X \ [{\rm keV}]$ & obsid(s) \\ \hline
2  &  0.310  &  $195.07 \pm 6.78 $  &  $200.65 \pm 6.90 $  &  $10.90^{+0.84}_{-0.81}$  &  9331,15099     \\ \hline
3  &  0.424  &  $174.46 \pm 5.07 $  &  $171.91 \pm 4.49 $  &  $7.39^{+0.41}_{-0.32}$  &  13396,16355,17536     \\ \hline
4  &  0.307  &  $146.24 \pm 4.04 $  &  $144.10 \pm 4.00 $  &  $10.24^{+0.26}_{-0.26}$  &  12260,16127,16282,16524,16525,16526     \\ \hline
5  &  0.355  &  $178.84 \pm 8.71 $  &  $188.40 \pm 10.06 $  &  $14.89^{+0.59}_{-0.55}$  &  4966     \\ \hline
6  &  0.373  &  $139.18 \pm 4.67 $  &  $138.85 \pm 4.72 $  &  $7.88^{+1.08}_{-0.80}$  &  13395     \\ \hline
8  &  0.243  &  $135.48 \pm 5.08 $  &  $136.44 \pm 4.69 $  &  $12.16^{+1.36}_{-0.92}$  &  15097     \\ \hline
10  &  0.330  &  $141.08 \pm 5.96 $  &  $142.26 \pm 6.28 $  &  $9.48^{+0.73}_{-0.53}$  &  11710,16285     \\ \hline
12  &  0.534  &  $160.33 \pm 6.45 $  &  $159.39 \pm 6.29 $  &  $7.51^{+3.04}_{-1.41}$  &  13466     \\ \hline
14  &  0.282  &  $129.00 \pm 4.30 $  &  $132.86 \pm 4.36 $  &  $9.46^{+0.66}_{-0.45}$  &  3248,11728     \\ \hline
15  &  0.610  &  $169.08 \pm 5.77 $  &  $165.92 \pm 5.63 $  &  $7.71^{+0.84}_{-0.55}$  &  12264,13116,13117     \\ \hline
16  &  0.289  &  $132.62 \pm 4.75 $  &  $130.37 \pm 4.73 $  &  $6.25^{+2.57}_{-1.30}$  &  17162,16271,17162     \\ \hline
17  &  0.597  &  $144.88 \pm 5.51 $  &  $152.04 \pm 5.00 $  &  $14.32^{+0.52}_{-0.52}$  &  13401,16135,16545     \\ \hline
19  &  0.421  &  $127.98 \pm 4.61 $  &  $124.99 \pm 4.26 $  &  $11.83^{+1.25}_{-0.90}$  &  12259     \\ \hline
20  &  0.231  &  $136.78 \pm 7.18 $  &  $135.36 \pm 6.69 $  &  $9.88^{+0.79}_{-0.66}$  &  15108     \\ \hline
21  &  0.350  &  $139.94 \pm 7.49 $  &  $125.67 \pm 5.83 $  &  $5.86^{+0.63}_{-0.37}$  &  17185     \\ \hline
\end{tabular}
\end{table*}

\begin{table*}
\caption{\xmm\ Clusters} \label{tab:XMM-table}
\begin{tabular}{|l|l|l|l|l|l|}
\hline
MEM\_MATCH\_ID & $z_{\lambda}$  & LAMBDA\_CHISQ & LAMBDA\_CHISQ (X-ray peak) & ${\rm k}T_X \ [{\rm keV}]$ & XCS\_NAME \\ \hline
1  &  0.430  &  $234.50 \pm 7.52 $  &  $238.88 \pm 7.37 $  &  $8.07^{+0.22}_{-0.21}$  &  XMMXCSJ025417.8-585705.2     \\ \hline
2  &  0.310  &  $195.07 \pm 6.78 $  &  $198.50 \pm 6.67 $  &  $6.11^{+0.14}_{-0.14}$  &  XMMXCSJ051636.6-543120.8     \\ \hline
3  &  0.424  &  $174.46 \pm 5.07 $  &  $171.91 \pm 4.79 $  &  $5.78^{+0.70}_{-0.59}$  &  XMMXCSJ041114.1-481910.9     \\ \hline
4  &  0.307  &  $146.24 \pm 4.04 $  &  $149.23 \pm 3.98 $  &  $8.46^{+0.25}_{-0.24}$  &  XMMXCSJ024529.3-530210.7     \\ \hline
5  &  0.355  &  $178.84 \pm 8.71 $  &  $190.51 \pm 10.17 $  &  $3.39^{+0.83}_{-0.56}$  &  XMMXCSJ224857.4-443013.6     \\ \hline
8  &  0.243  &  $135.48 \pm 5.08 $  &  $135.74 \pm 4.70 $  &  $9.48^{+0.36}_{-0.34}$  &  XMMXCSJ213516.8-012600.0     \\ \hline
10  &  0.330  &  $141.08 \pm 5.96 $  &  $140.76 \pm 5.94 $  &  $7.06^{+0.78}_{-0.67}$  &  XMMXCSJ213511.8-010258.0     \\ \hline
14  &  0.282  &  $129.00 \pm 4.30 $  &  $134.66 \pm 4.43 $  &  $7.45^{+0.24}_{-0.22}$  &  XMMXCSJ233738.6-001614.5     \\ \hline
15  &  0.610  &  $169.08 \pm 5.77 $  &  $164.19 \pm 5.68 $  &  $7.29^{+0.76}_{-0.62}$  &  XMMXCSJ055943.5-524937.5     \\ \hline
17  &  0.597  &  $144.88 \pm 5.51 $  &  $150.54 \pm 4.93 $  &  $12.11^{+0.07}_{-0.07}$  &  XMMXCSJ234444.0-424314.2     \\ \hline
19  &  0.421  &  $127.98 \pm 4.61 $  &  $125.71 \pm 4.58 $  &  $8.31^{+0.24}_{-0.23}$  &  XMMXCSJ043818.3-541916.5     \\ \hline
20  &  0.231  &  $136.78 \pm 7.18 $  &  $134.34 \pm 6.71 $  &  $8.05^{+0.48}_{-0.44}$  &  XMMXCSJ202323.2-553504.7     \\ \hline
24  &  0.494  &  $126.99 \pm 4.31 $  &  $127.26 \pm 4.33 $  &  $5.79^{+0.17}_{-0.16}$  &  XMMXCSJ024339.4-483338.3     \\ \hline
25  &  0.427  &  $130.39 \pm 6.17 $  &  $131.88 \pm 6.31 $  &  $5.26^{+0.71}_{-0.56}$  &  XMMXCSJ213538.5-572616.6     \\ \hline
26  &  0.450  &  $138.53 \pm 6.45 $  &  $138.08 \pm 6.31 $  &  $6.41^{+0.28}_{-0.26}$  &  XMMXCSJ030415.7-440153.0     \\ \hline
\end{tabular}
\end{table*}

\newpage

\section{X-ray emission peak centering sensitivity} \label{app:centering-sensitivity}

\begin{figure}
    \centering
    \includegraphics[width=0.48\textwidth]{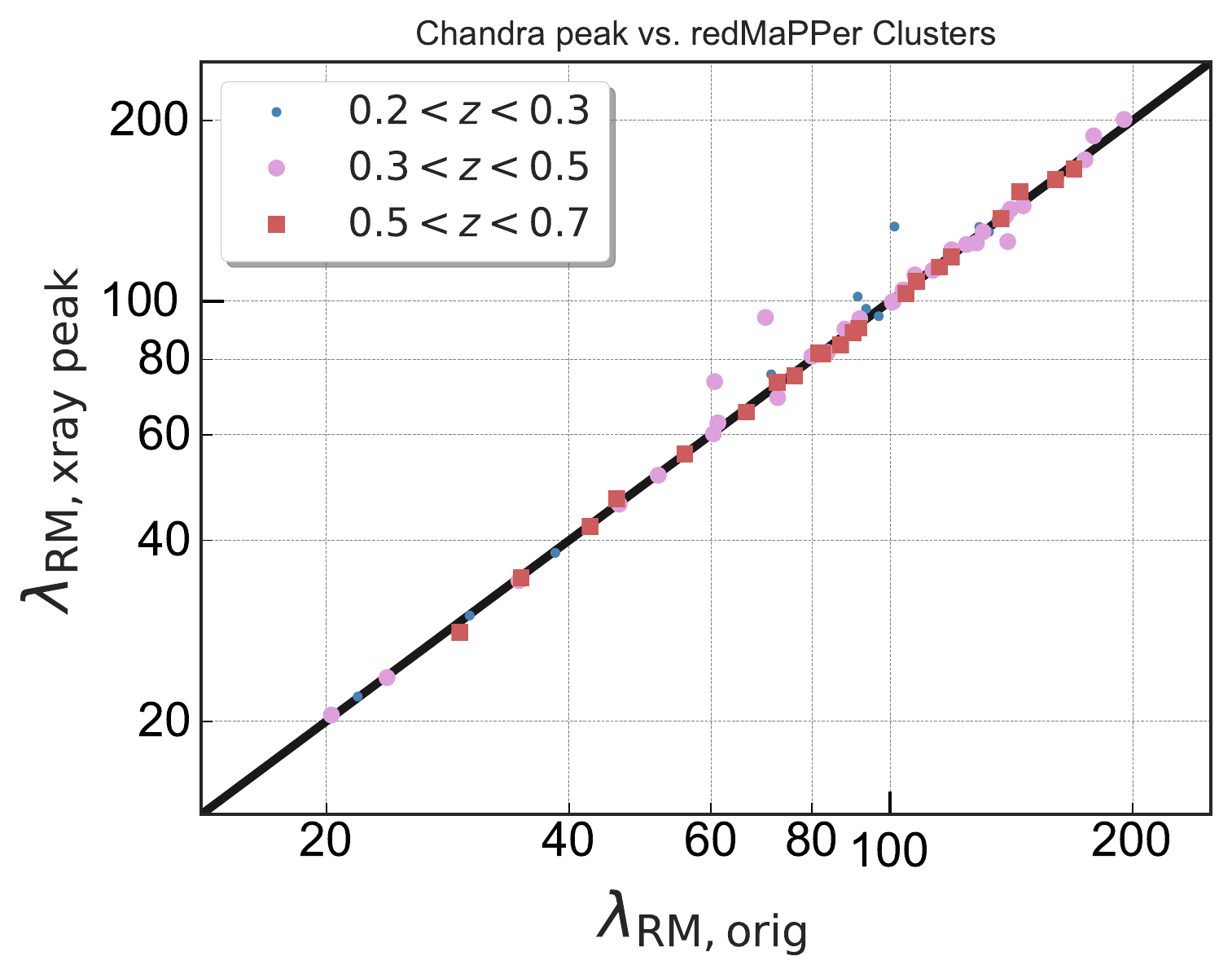}
    \includegraphics[width=0.48\textwidth]{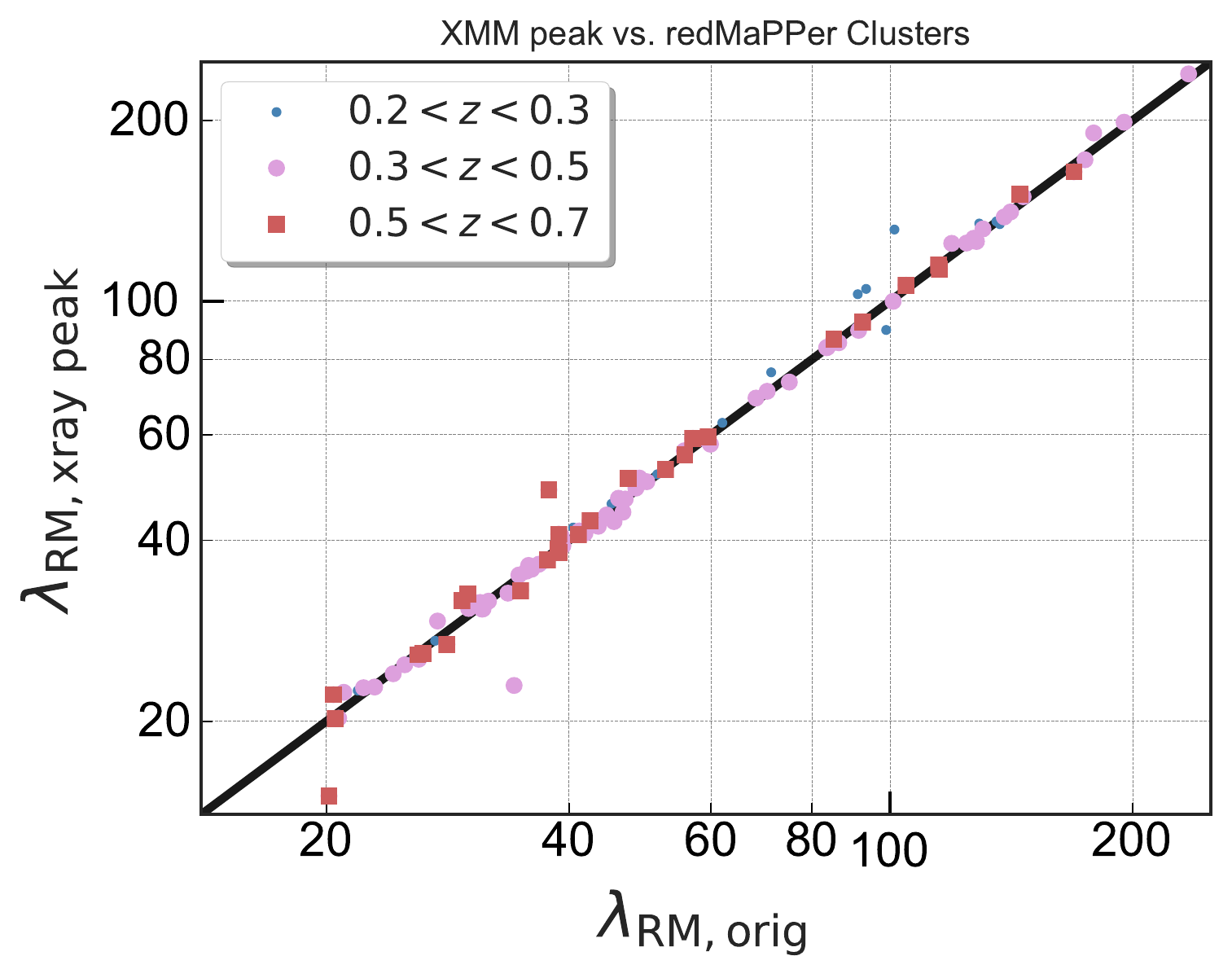}
    \caption{ Re-estimation of \RM-selected clusters richness at the X-ray emission peak versus their original \RM\ richness. Left (right) panel shows the clusters with a matched X-ray source in the Chandra (XMM) archival dataset. The black line is a unity line for the reference. }
    \label{fig:richness-comp}
\end{figure}

We are concerned about the relation between the properties of the \RM-selected cluster observables and its host halo. Therefore, we need to correct for the fraction of the mis-centered population. The mass scale of the \RM\ host halos is studied in \citet{DESY1-WL:2018} by correcting for the mis-centered clusters. Instead of modeling, we correct our cluster observables with an associated X-ray center. 
To assign a center, we assume the hot gas content of galaxy clusters traces the gravitational potential sourced by the host halo. Specifically, we estimate the center of the host halo with the location of the X-ray emission peak. We run the \RM\ algorithm and assign a new optical richness to each X-ray extended source which is matched to a \RM-selected cluster. Figure \ref{fig:richness-comp} shows the assigned richness at the X-ray emission peak versus the original \RM\ richness. Except for a handful of mis-centered clusters, the change in the richness is negligible \citep[see][for more detail]{Zhang:2018centering}. We also find that the estimated richness is insensitive to the data obtained by different instruments and the two X-ray analysis pipelines. 

\section{Richness--temperature covariance} \label{app:correlation}

\begin{figure}
    \centering
    \includegraphics[width=0.48\textwidth]{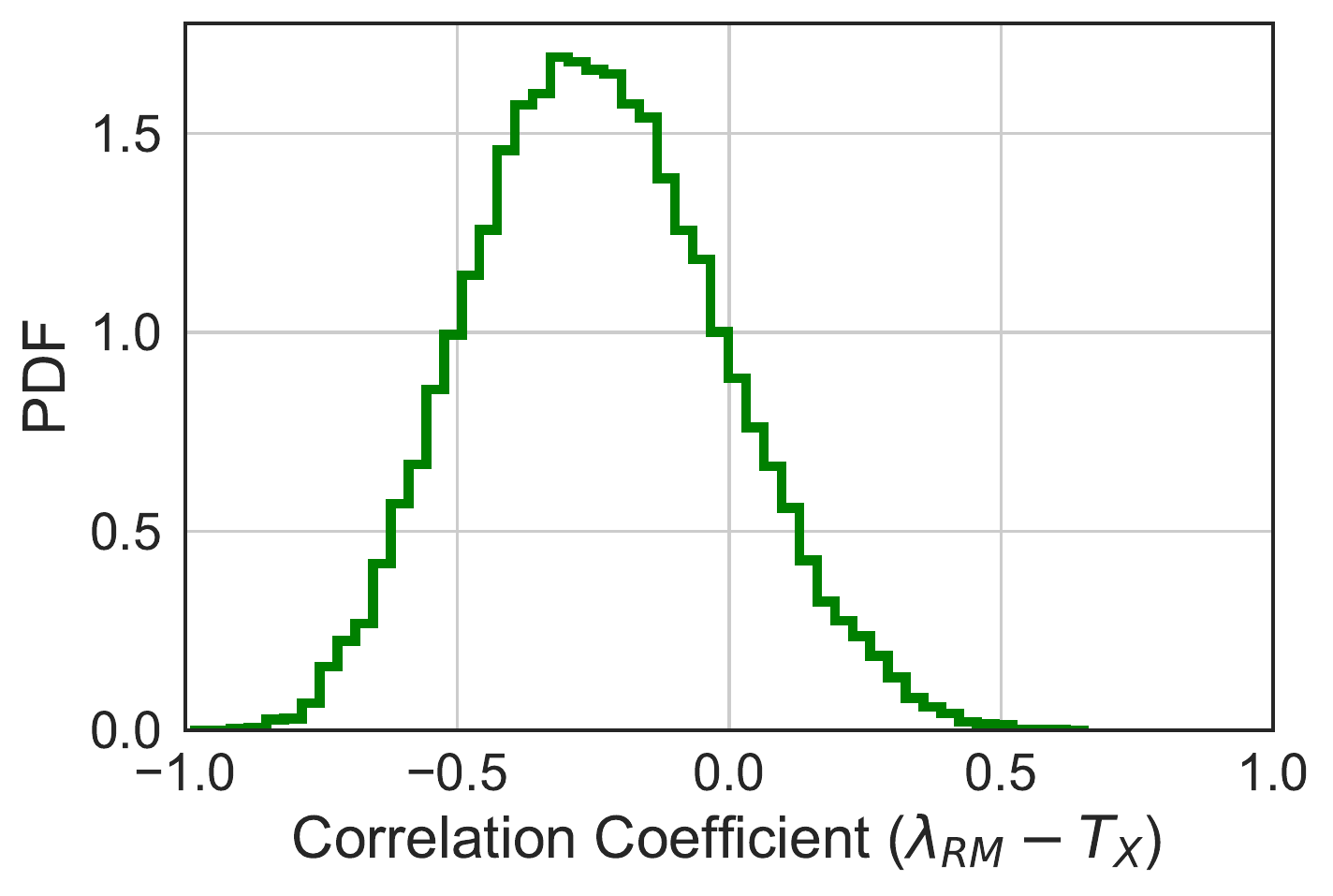}
    \caption{The marginalized posterior distribution for the $\ln(\lamRM)$ and $\ln(T_X)$ correlation coefficient about fixed host halo mass employed in this work. This is taken from Farahi et al. (in preparation).}
    \label{fig:prior-r}
\end{figure}

Farahi et al. (in preparation) studied the full property covariance of ten observables, including \RM\ richness and X-ray temperature, regressed on the weak-lensing mass. Their sample consist of a 41 X-ray luminosity selected, low-redshift clusters with a weak-lensing mass measurement for each individual cluster. Figure~\ref{fig:prior-r} presents their marginalized posterior distribution for the $\ln(\lamRM)$ and $\ln(T_X)$ correlation coefficient about weak-lensing mass, which is employed in this work. Clearly a strong positive and negative correlations are ruled out with a high statistical significance. 

\section{Running of variance with richness}\label{app:running}

\begin{figure}
      \centering
      \includegraphics[width=0.49\textwidth]{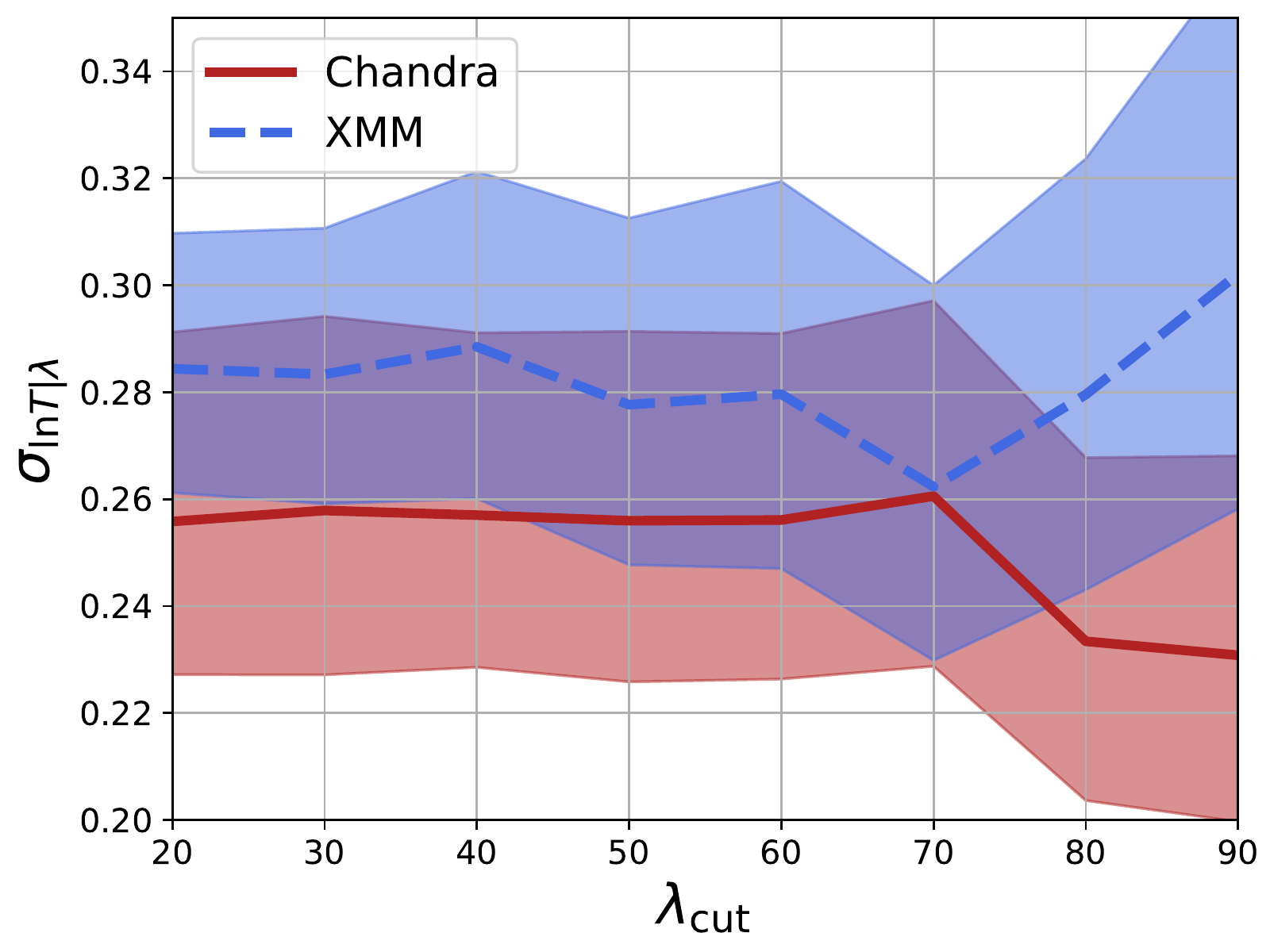}
      \caption{Constraints on the scatter of the $T_X$--$\lamRM$ relation derived from subsamples of \chandra\ (blue, dashed line) and \xmm\ (red solid line) data thresholded by \RM\ richness, $\lamRM > \lambda_{\rm cut}$. Shaded regions show $68\%$ confidence intervals. }
      \label{fig:selection-sig}
  \end{figure}

We further study the change in the scatter parameter for a subset of clusters by progressively applying $\lamRM > \lambda_{\rm cut}$ (Fig. \ref{fig:selection-sig}). We find that within the $68\%$ statistical confidence intervals the estimated intrinsic scatter about the mean relation does not change. This implies that the bias caused by the X-ray analysis pipeline is negligible, or otherwise there is a miraculous running of the scatter that cancels the X-ray selection bias. Saying that, one should be cautious that a different subset of \RM\ cluster sample, with a larger sample size or a different X-ray analysis pipeline, can have different characteristics.

\label{lastpage}

\end{document}